\documentclass[acmsmall]{acmart}
\AtBeginDocument{%
  }

\setcopyright{acmlicensed}
\copyrightyear{2026}
\acmYear{2026}
\acmDOI{XXXXXXX.XXXXXXX}
\acmJournal{TACO}

\usepackage{subcaption}
\usepackage{amsmath}
\usepackage{tikz}
\usepackage[linesnumbered,ruled,vlined,noend]{algorithm2e}
\SetAlgoSkip{}
\SetCommentSty{mycommfont}

\usepackage[most]{tcolorbox}

\newcommand{\tb}[1]{\textbf{#1}}

\newcommand{\name}{FLAMENCO}

\newcommand{\squishlist}{
   \begin{list}{$\bullet$}
    { 
    \setlength{\itemsep}{0pt}      \setlength{\parsep}{0pt}
      \setlength{\topsep}{3pt}       \setlength{\partopsep}{0pt}
      \setlength{\listparindent}{-2pt}
      \setlength{\itemindent}{-5pt}
      \setlength{\leftmargin}{1em} \setlength{\labelwidth}{0em}
      \setlength{\labelsep}{0.5em} } }

\newcommand{\squishend}{
    \end{list}  }

\newcommand*\circled[1]{\tikz[baseline=(char.base)]{
  \node[shape=circle,draw,fill=black,text=white,font=\bf,inner sep=0.5pt] (char)
  {\scriptsize#1};
}}

\newcommand{\etal}{et al.}
\newcommand{\ie}{i.e.}

\newcommand{\etc}{etc.}

\newcommand{\COMMENT}[1]{#1}

\renewcommand{\COMMENT}[1]{}

\begin{document}

\title{A System Architecture for Low Latency Multiprogramming Quantum Computing}

\author{Yilun Zhao}
\email{zyilun8@gmail.com}
\orcid{0000-0002-6812-5120}
\affiliation{%
  \institution{State Key Lab of Processors, Institute of Computing Technology, Chinese Academy of Sciences}
  \city{Beijing}
  \state{Beijing}
  \country{China}
}
\affiliation{%
  \institution{University of Chinese Academy of Sciences}
  \city{Beijing}
  \state{Beijing}
  \country{China}
}

\author{Yu Chen}
\email{chenyu21b@ict.ac.cn}
\orcid{0000-0001-9279-9515}
\affiliation{%
  \institution{State Key Lab of Processors, Institute of Computing Technology, Chinese Academy of Sciences}
  \city{Beijing}
  \state{Beijing}
  \country{China}
}
\affiliation{%
  \institution{University of Chinese Academy of Sciences}
  \city{Beijing}
  \state{Beijing}
  \country{China}
}

\author{Kaiyan Chang}
\email{changkaiyan@live.com}
\orcid{0000-0003-1920-0101}
\affiliation{%
  \institution{State Key Lab of Processors, Institute of Computing Technology, Chinese Academy of Sciences}
  \city{Beijing}
  \state{Beijing}
  \country{China}
}
\affiliation{%
  \institution{University of Chinese Academy of Sciences}
  \city{Beijing}
  \state{Beijing}
  \country{China}
}

\author{He Li}
\email{helix@seu.edu.cn}
\orcid{0000-0001-8961-4487}
\affiliation{%
  \institution{Southeast University}
  \city{Nanjing}
  \state{Jiangsu}
  \country{China}
}

\author{Bing Li}
\email{libing9084@gmail.com}
\orcid{0000-0003-0732-2267}
\affiliation{%
  \institution{Institute of Microelectronics, Chinese Academy of Sciences}
  \city{Beijing}
  \state{Beijing}
  \country{China}
}

\author{Yinhe Han}
\email{yinhes@ict.ac.cn}
\orcid{0000-0003-0904-6681}
\affiliation{%
  \institution{State Key Lab of Processors, Institute of Computing Technology, Chinese Academy of Sciences}
  \city{Beijing}
  \state{Beijing}
  \country{China}
}

\author{Ying Wang}
\email{wangying2009@ict.ac.cn}
\orcid{0000-0001-5172-4736}
\affiliation{%
  \institution{State Key Lab of Processors, Institute of Computing Technology, Chinese Academy of Sciences}
  \city{Beijing}
  \state{Beijing}
  \country{China}
}

\renewcommand{\shortauthors}{Y. Zhao et al.}

\begin{abstract}

As quantum systems scale, multiprogramming quantum computing (MPQC) provides a practical way to improve device utilization and throughput. However, because quantum executables are device-dependent, non-portable across qubit regions, and highly susceptible to noise and crosstalk, current MPQC pipelines rely on expensive online compilation to co-optimize concurrently running programs. This online step dominates runtime and impedes low-latency deployments for practical, real-world workloads in the future, such as repeatedly invoked quantum neural network (QNN) services.

We present FLAMENCO, a fidelity-aware multi-version compilation system that enables independent offline compilation and low-latency multiprogramming at runtime. \tb{At the architecture level}, the system abstracts devices into compute units to reduce the search space of region allocation. \tb{At compile time}, it generates diverse executable versions for each program---each bound to a distinct qubit region---allowing dynamic region selection at runtime and overcoming non-portability. \tb{At runtime}, it employs a lightweight orchestrator that uses post-compilation fidelity metrics to avoid conflicts and mitigate crosstalk, supporting conflict-free co-execution without online co-optimization. Evaluations show that FLAMENCO achieves over 5$\times$ runtime speedup in post-scheduling execution while maintaining comparable execution fidelity on common-success workloads. When integrated into existing scheduler-coupled systems, it raises workload-level conflict-free orchestration ratio from 0.183 to 1.000 for HyperQ and from 0.050 to 0.400 for QOS.

\end{abstract}



\maketitle

\section{Introduction}

\emph{Quantum computing} (QC) has undergone rapid evolution.
The number of quantum bits (\emph{qubits}) in a quantum device has increased from dozens to hundreds during the past decade~\cite{monz14QubitEntanglementCreation2011,acharyaQuantumErrorCorrection2025,bluvsteinLogicalQuantumProcessor2023}.
As quantum devices scale, recent studies have demonstrated the potential of QC to solve real-world problems in application scenarios, such as finance~\cite{hermanQuantumComputingFinance2023}, physics~\cite{kimEvidenceUtilityQuantum2023,abaninObservationConstructiveInterference2025}, and machine learning~\cite{jiangCodesignFrameworkNeural2021}.
These advances have fueled research and commercial interests, resulting in increasing demand to access quantum computers.

To increase system throughput and device utilization, multiprogramming quantum computing (MPQC) has been proposed to execute multiple programs simultaneously on the same system~\cite{dasCaseMultiProgrammingQuantum2019,liuQuCloudNewQubit2021}.
In contrast to its classical counterpart~\cite{dijkstraStructureTHEmultiprogrammingSystem1968a,krauseTaskschedulingAlgorithmMultiprogramming1973}, quantum devices present unique challenges to the implementation of multiprogramming due to intrinsic fragility caused by quantum mechanical properties, including short qubit lifetimes, high gate error rates, and crosstalk~\cite{patelQUESTSystematicallyApproximating2022,muraliNoiseAdaptiveCompilerMappings2019,xieSuppressingZZCrosstalk2022,dingSystematicCrosstalkMitigation2020,tannuHAMMERBoostingFidelity2022}.
As a consequence, present MPQC design requires \emph{online\footnote{In this paper, ``online'' refers to events or phases that occur at runtime; ``offline'' refers to events or phases that occur before runtime.} compilation} to co-optimize all quantum programs at runtime to achieve a conflict-free execution with high fidelity ($\S$~\ref{ssec:quantum_multiprogramming}).

However, online quantum compilation introduces high runtime overhead because it performs device-dependent optimizations to overcome hardware limitations~\cite{liTacklingQubitMapping2019,muraliNoiseAdaptiveCompilerMappings2019,siraichiQubitAllocation2018,patelQUESTSystematicallyApproximating2022} ($\S$~\ref{ssec:quantum_compilation}). This challenge applies not only to today's noisy intermediate-scale quantum (NISQ) devices but also to future fault-tolerant quantum computers (FTQC).
If quantum computers become widely used computing infrastructure, low-latency execution will be important.
This latency requirement creates a mismatch with existing MPQC designs: they co-compile the selected programs at runtime, placing expensive compilation on the per-invocation critical path.
Consider quantum neural network (QNN) models~\cite{wangQuantumNASNoiseAdaptiveSearch2022,wangSymmetricPruningQuantum2023} deployed as repeated-invocation quantum services. Their circuit structures can remain stable across invocations, akin to classical machine-learning services, so spending most of the co-execution runtime on compilation is inefficient.
\begin{figure}[t!]
    \begin{subfigure}{0.75\textwidth}
        \centering
        \includegraphics[width=\linewidth]{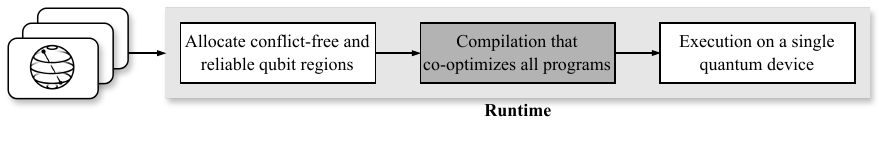}
        \caption{Workflow of existing MPQC.}
        \label{fig:workflow_mqc}
    \end{subfigure}

    \begin{subfigure}{0.75\textwidth}
        \centering
        \includegraphics[width=\linewidth]{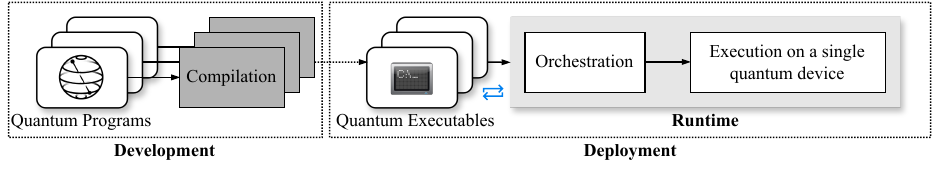}
        \caption{Target workflow of this work.}
        \label{fig:workflow_qvm}
    \end{subfigure}
    \caption{(a) Current MPQC requires region allocation before compilation, which places expensive online compilation on the critical path. (b) \name{} targets a development/deployment/invocation lifecycle: programs are compiled into reusable executable-version libraries during development, deployed with the service, and orchestrated at each post-deployment invocation to resolve conflicts and preserve fidelity.}
    \Description{Two workflow diagrams. The existing multiprogramming workflow allocates regions and compiles programs after a co-execution set is selected. The FLAMENCO workflow compiles reusable physical-region executable versions during development and selects conflict-free versions at each invocation.}
    \label{fig:qvm_motivation_workflow}
\end{figure}

In this paper, we identify the root cause of online compilation in MPQC: the requirement to statically allocate conflict-free qubit regions prior to executable generation (Figure~\ref{fig:workflow_mqc}).
To overcome this obstacle, \name{} moves compilation to the development stage before deployment: it compiles each program into multiple physical-region executable versions offline, deploys the resulting executable library, and then dynamically \emph{orchestrates} these versions during repeated post-deployment invocations to achieve conflict-free, high-fidelity co-execution (Figure~\ref{fig:workflow_qvm}).
This development/deployment/invocation split is common in classical computing, but it creates two quantum-specific challenges.
First, the runtime must avoid region conflicts after programs have been compiled into physical-region executables.
This is an intrinsic challenge in QC since quantum executables are tightly coupled with the backend physical qubits, \ie, a quantum executable compiled on a specific qubit region cannot run on another region correctly.
Second, the system must preserve execution fidelity when programs are compiled independently.
Although MPQC inevitably compromises the fidelity of individual programs~\cite{dasCaseMultiProgrammingQuantum2019}, online compilation allows for co-optimizations among all programs to mitigate this issue of fidelity loss.
Offline compilation removes these inter-program co-optimization opportunities from the runtime path, making fidelity-aware orchestration necessary.
Together, these two challenges require a mechanism that separates offline compilation from per-invocation orchestration.

To this end, we design a \tb{F}ide\tb{L}ity-\tb{A}ware \tb{M}ulti-v\tb{E}rsio\tb{N} \tb{CO}mpliation system, named \name{}.
The system addresses these challenges through three design components.
(i) At the architecture level, we abstract quantum devices into compute units and allocate qubits at compute-unit granularity, which reduces the region-allocation search space ($\S$~\ref{ssec:arch_abstraction}).
(ii) At compile time, we generate multiple executable versions for each program, each bound to a distinct qubit region. This design defers concrete region choice to runtime and overcomes the non-portable nature of quantum executables ($\S$~\ref{ssec:multi_version_compilation}). The compiler also estimates executable fidelity to support runtime selection.
(iii) At runtime, we use a lightweight heuristic to orchestrate executable versions with fidelity awareness ($\S$~\ref{ssec:fid_aware_orch}).
Together, these components reduce online compilation latency while preserving execution fidelity, and in some cases improve it relative to existing online compilation schemes.
We also study multiple orchestration strategies and their scaling behavior as the number of concurrent programs increases ($\S$~\ref{ssec:scalability}).

In sum, our main contributions include:

\squishlist{}
    \item \textbf{A Novel Perspective in MPQC.}
    We identify a central bottleneck in existing MPQC designs: time-intensive online compilation can dominate quantum-program runtime.
    We argue that future low-latency MPQC should remove online compilation from the per-invocation critical path.
    \item \textbf{\name{}}.
    We design and implement a system architecture for future low-latency MPQC.
    Key innovations include:
    (i) a multi-version compilation strategy enabled by a clear architectural abstraction of quantum devices into compute units, and
    (ii) a simple-yet-effective fidelity-aware orchestration mechanism at runtime.
    \item \textbf{Systematic Evaluation.}
    We evaluate latency, fidelity, and utilization across noisy simulators, real quantum devices, state-of-the-art MPQC baselines, and recent quantum-cloud systems such as HyperQ and QOS.
    Results show over 5$\times$ runtime speedup in post-scheduling execution and raise workload-level conflict-free orchestration ratio from 0.183 to 1.000 for HyperQ~\cite{taoQuantumVirtualMachines2025} and from 0.050 to 0.400 for QOS~\cite{giortamis2025qosquantumoperating}, while maintaining comparable common-success execution fidelity.

\squishend{}

\section{Background and Motivation}

In this section, we offer a concise overview of QC.
QC research spans multiple communities, including quantum algorithms, computer architecture, compilation, control systems, and device physics~\cite{nielsenQuantumComputationQuantum2010,muraliFullstackRealsystemQuantum2019,krantzQuantumEngineersGuide2019}.
Our discussion will focus on providing the essential information needed to comprehend the current design of MPQC and its constraints.

\subsection{Quantum Basics}
\label{ssec:quantum_basic_and_qc_cloud}

The capability of QC arises from the utilization of qubits as fundamental units.
A qubit can exist in a \emph{superposition} state $\left|\psi\right> = \alpha \left|0\right> + \beta \left|1\right>$, where $\alpha$ and $\beta$ are complex amplitudes.
Quantum operations are responsible for manipulating qubit states.
Of particular significance are entangling gates, including the controlled-NOT (CNOT) gate, which can entangle qubits.
Roughly speaking, superposition and entanglement allow quantum computers to evaluate a function $f(x)$ for many different values simultaneously, offering extensive \emph{quantum parallelism}~\cite{nielsenQuantumComputationQuantum2010},
serving as the backbone of quantum algorithms~\cite{shorPolynomialTimeAlgorithmsPrime1999,deutsch1985quantum}.
These algorithms are implemented by \emph{quantum circuits}\footnote{This paper focuses on circuit-level workloads. We use \emph{quantum circuit} and \emph{quantum program} interchangeably when no classical branching or control-hardware computation is involved.},
which essentially comprise a sequence of quantum operations.
A typical quantum circuit initializes qubits to superposition states, performs task-specific operations, and conducts \emph{measurements} that collapse qubits to deterministic states at the end.

\subsection{Quantum Compilation}

\label{ssec:quantum_compilation}

\begin{figure}[t!]
    \centering
    \captionsetup[subfigure]{belowskip=0pt,aboveskip=-2pt}
    \begin{subfigure}[t]{0.36\textwidth}
        \centering
        \includegraphics[width=\linewidth]{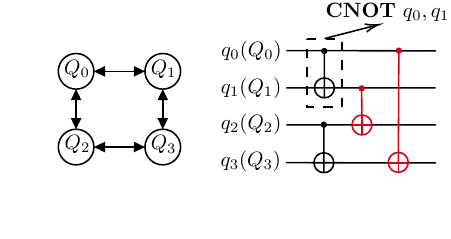}
        \caption{}
        \label{fig:sample_mapping}
    \end{subfigure}\hfill
    \begin{subfigure}[t]{0.4\textwidth}
        \includegraphics[width=\linewidth]{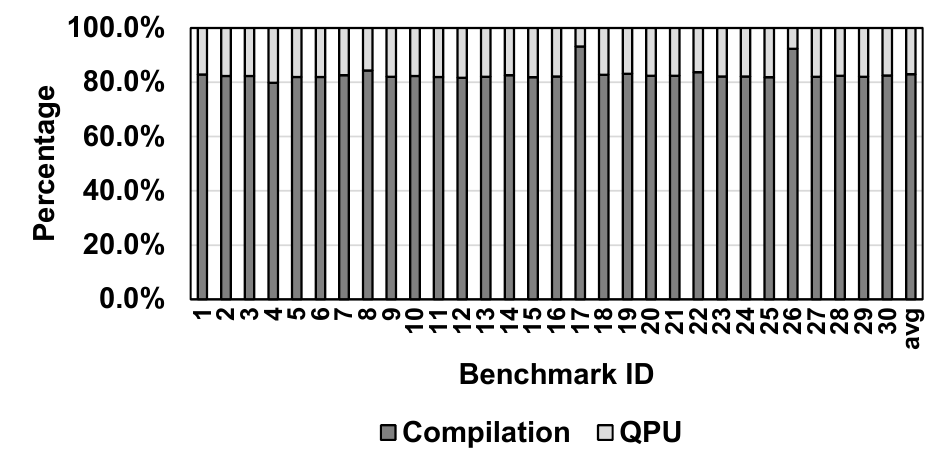}
        \caption{}
        \label{fig:compilation_vs_duration}
    \end{subfigure}\hfill
    \begin{subfigure}[t]{0.19\textwidth}
        \includegraphics[width=\linewidth]{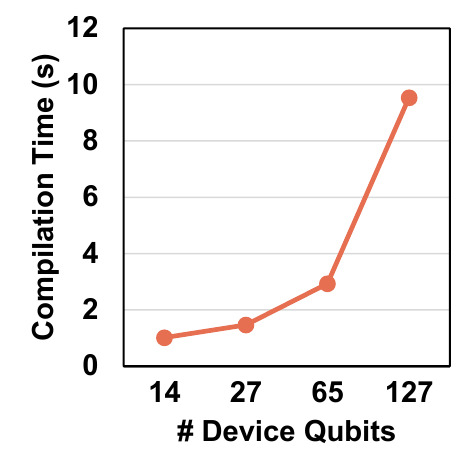}
        \caption{}
        \label{fig:compilation_vs_qubits}
    \end{subfigure}
	\caption{Quantum compilation and its time cost in MPQC. (a) A toy example of qubit mapping problem~\cite{liTacklingQubitMapping2019}; (b) Breakdown of execution time. Each benchmark is denoted by its \texttt{ID} as listed in Table~\ref{tab:benchmarks}; (c) Compilation time vs. \# qubits.}
    \Description{Three panels illustrate qubit mapping and compilation cost. The first maps a logical circuit to connected physical qubits. The second breaks execution time into compilation and QPU time for the evaluated benchmarks. The third shows compilation time increasing with circuit qubit count.}
\end{figure}

In this paper, quantum compilation refers to passes that transform a circuit into an executable form for a target backend or qubit region, such as mapping, routing, and nativization.
Here we omit some steps such as nativization~\cite{dasImitationGameLeveraging2023} and synthesis~\cite{patelQUESTSystematicallyApproximating2022}, focusing on \emph{qubit mapping and routing problem}~\cite{liTacklingQubitMapping2019}, a key bottleneck for limited-connectivity NISQ devices.
This problem arises as implementing a two-qubit gate on many real quantum devices requires a physical connection, which means that two-qubit gates can only be applied on two physically adjacent qubits.
However, quantum devices, e.g., the IBM superconducting devices~\cite{IBM_heavy_hex}, typically have limited connectivity.
Figure~\ref{fig:sample_mapping} shows a 4-qubit device model, where the bidirectional arrows denote two-qubit connections.
In the circuit shown in Figure~\ref{fig:sample_mapping}, the first two CNOT gates can be executed directly, while the later two CNOT gates (marked red) cannot be executed because the corresponding qubit pairs ($\{Q_1,Q_2\}$ and $\{Q_0,Q_3\}$) are not connected on the device.
To solve this problem, we can employ SWAP operations, which can be implemented using additional two-qubit gates, to exchange the states between two qubits.
For instance, we can insert a SWAP gate between $q_0$ and $q_1$ before executing the last two two-qubit gates.
Consequently, the last two two-qubit gates can operate on adjacent physical qubit pairs $\{Q_0,Q_2\}$ and $\{Q_1,Q_3\}$.
However, adding extra two-qubit gates is expensive, especially in current NISQ~\cite{preskillQuantumComputingNISQ2018} devices.
Given that qubits maintain their state for only a short period (a few hundreds of microseconds), the increased circuit depth and extended execution time from additional two-qubit gates elevate the likelihood of decoherence errors~\cite{krantzQuantumEngineersGuide2019}.
Therefore, finding the optimal mapping and minimizing the additional SWAP operations~\cite{liTacklingQubitMapping2019,zhangTimeoptimalQubitMapping2021} becomes an important objective for compilers targeting limited-connectivity devices.

This process has two notable characteristics.

\squishlist{}
    \item \textbf{Time-intensive}. Finding the optimal solution is an NP-complete problem~\cite{siraichiQubitAllocation2018}.
    Solver-based solutions~\cite{molaviQubitMappingRouting2022,tanOptimalLayoutSynthesis2020} result in smaller circuit depth but suffer from long runtime.
    While heuristic approaches trade off circuit depth for better efficiency, the runtime is still long compared to the circuit execution time on quantum devices, as we show later in $\S$~\ref{ssec:problem_and_motivation}.
    \item \textbf{Non-portable}.
    When mapped to a qubit region, the SWAP operation placement is affected by the region's topology.
    Additionally, on machines where calibrated two-qubit gate error rates vary across links, compilers can also account for such variation.
    These factors are taken into account during compilation~\cite{muraliNoiseAdaptiveCompilerMappings2019}.
    Consequently, the compilation outcomes are determined by both connectivity and error rates.
    Given that these factors vary across regions, a quantum executable compiled for one region may not work correctly on another.

\squishend{}

\begin{tcolorbox}[colback=gray!10!white,colframe=gray!10!black,sharp corners,left=0mm,right=0mm,top=0mm,bottom=0mm,title=Characteristics of Quantum Compilation in NISQ and FTQC]

Although the preceding discussion focuses on quantum compilation for NISQ devices, similar challenges can persist for future large-scale FTQC.
For instance, mapping quantum low-density parity-check (QLDPC) codes onto superconducting processors continues to incur substantial routing overhead in both gate count and circuit depth, due to the inherent mismatch between the codes’ frequent long-range interactions and the hardware’s fixed, local connectivity~\cite{yangQubitMappingRouting2025}.
More generally, some QEC layouts can require interactions that are difficult to realize directly on sparse or irregular hardware connectivity, although the severity of this mismatch depends on the code and hardware modality.
This disparity is further exacerbated by non-uniformly distributed defective qubits, introducing additional complexity to the mapping problem and potentially giving rise to new optimization challenges~\cite{yinQECCSynthLayoutSynthesizer2025}.
Consequently, time-consuming compilation and limited portability are likely to remain intrinsic features of quantum compilation in the long term.

\end{tcolorbox}

\subsection{Multiprogramming Quantum Computers}

\label{ssec:quantum_multiprogramming}

The concept of MPQC is proposed by Das \etal~\cite{dasCaseMultiProgrammingQuantum2019} to improve system throughput,
which extends the mapping and routing problem to multiple quantum programs, consisting of the following two key steps.

\squishlist{}
    \item \textbf{Qubit Allocation.} Executing a quantum program with $k$ qubits necessitates a region of $k$ connected physical qubits.
    The objective of qubit allocation is to designate a specific qubit region for each program while preventing conflicts.
    Crucially, existing approaches employ both program and device characteristics to develop an allocation scheme that balances all programs, considering both fairness and fidelity~\cite{dasCaseMultiProgrammingQuantum2019,liuQuCloudNewQubit2021,niuHowParallelCircuit2022}.
    \item \textbf{Ensembled Compilation and Execution.} Once the qubit regions are determined, the co-running programs are collectively transformed into executable formats.
    To improve execution fidelity, inter-program optimizations can be implemented on neighbouring regions.
    For instance, inserting inter-program SWAPs~\cite{liuQuCloudNewQubit2021} helps reduce the overall depth of the programs.
\squishend{}

\noindent
The above workflow of MPQC implies several unique properties.

\squishlist{}

\item \textbf{Spatial Sharing.}
Traditional classical multiprogramming enables processors to be temporally shared by multiple programs~\cite{dijkstraStructureTHEmultiprogrammingSystem1968a,millsMultiprogrammingSmallsystemsEnvironment1969,tuckerProcessControlScheduling1989}.
However, temporal sharing is hard to implement for NISQ-style QC since the states of qubits cannot be copied\footnote{Due to the quantum no-cloning theorem~\cite{nielsenQuantumComputationQuantum2010}.} and the lifetimes of qubits are extremely short.
Therefore, multiple quantum programs non-preemptively run on different qubit regions without interruption.

\item \textbf{Lock-step Execution.}
In a quantum device, multiple qubits may share the same readout line~\cite{zettles262DesignConsiderations2022}, which means multiple qubits need to be measured together.
To prevent the measurement of one program from interfering with the states of other programs, MPQC essentially combines multiple programs into a single one to run in lock-step~\cite{dasCaseMultiProgrammingQuantum2019,liuQuCloudNewQubit2021}.

\item \textbf{Online Compilation.}
Unlike classical multiprocessor systems, where a pre-compiled executable can be easily scheduled to run on different processors~\cite{tuckerProcessControlScheduling1989,gamsaTornadoMaximizingLocality1999a},
MPQC requires compiling multiple programs at runtime to avoid resource conflicts due to the non-portable property of quantum compilation.

\squishend{}

\subsection{Problem and Motivation }
\label{ssec:problem_and_motivation}

In this section, we profile the latency ($T_{\texttt{total}}$) of current MPQC design and identify that \tb{online compilation becomes the bottleneck.}
In many cloud-based quantum computing systems, a quantum program transitions through \emph{queued}, \emph{running}, and \emph{completed} states after submission.
Queue scheduling is orthogonal to \name{}: a queue scheduler decides which programs should co-run, whereas \name{} orchestrates the selected co-running set.
In this paper, \emph{latency} specifically denotes the duration of the running state, i.e., the time from the start to the end of program execution, excluding any queuing or scheduling delays.
As illustrated in $\S$~\ref{ssec:quantum_multiprogramming}, the latency of current MPQC includes online compilation time and the time spent on a quantum processing unit (QPU), i.e., the combination of the control electronics and quantum devices.
The control electronics execute instructions to apply device-level control signals, such as microwave pulses on superconducting devices, to control and readout state of qubits~\cite{zettles262DesignConsiderations2022}.
Hence, the latency is calculated as $T_{\texttt{total}} = T_{\texttt{compile}} + T_{\texttt{QPU}}$.
The compilation time ($T_{\texttt{compile}}$) is directly measured during the compilation process using IBM Qiskit~\cite{javadi-abhariQuantumComputingQiskit2024}.
Here, $T_{\texttt{QPU}}$ refers to the instruction-execution time of the quantum control processor that drives the QPU stack. At a fundamental level, the program runtime is governed by the control processor's instruction stream execution time.
Since the accurate QPU time ($T_{\texttt{QPU}}$) was not directly available from our IBM cloud measurements, we estimate it using an in-house, cycle-accurate, experimentally validated quantum control architecture simulator (CACTUS)~\cite{cactus,zhaoDistributedHISQDistributedQuantum2025a}, and then multiply the per-shot runtime by 1024 shots.
For the CACTUS timing parameters in this evaluation, we set 20 ns (40 ns) for single- (two-) qubit gates and 300 ns for measurements.

Figure~\ref{fig:compilation_vs_duration} shows that compilation is a significant performance bottleneck even with the widely used SABRE~\cite{liTacklingQubitMapping2019} algorithm from Qiskit.
More intricate optimizations, such as circuit synthesis~\cite{patelQUESTSystematicallyApproximating2022} or solver-based qubit mapping and routing~\cite{tanOptimalLayoutSynthesis2020,molaviQubitMappingRouting2022}, would likely increase this gap, and Figure~\ref{fig:compilation_vs_qubits} further shows that compilation time escalates with device scale.
Because QPU time for a fixed circuit and shot configuration is governed by physical gate durations, readout time, and qubit reinitialization latency~\cite{fuExperimentalMicroarchitectureSuperconducting2017}, low-latency MPQC systems should remove online compilation from the critical execution path.

\begin{figure*}[h!]
    \centering
    \includegraphics[width=\textwidth]{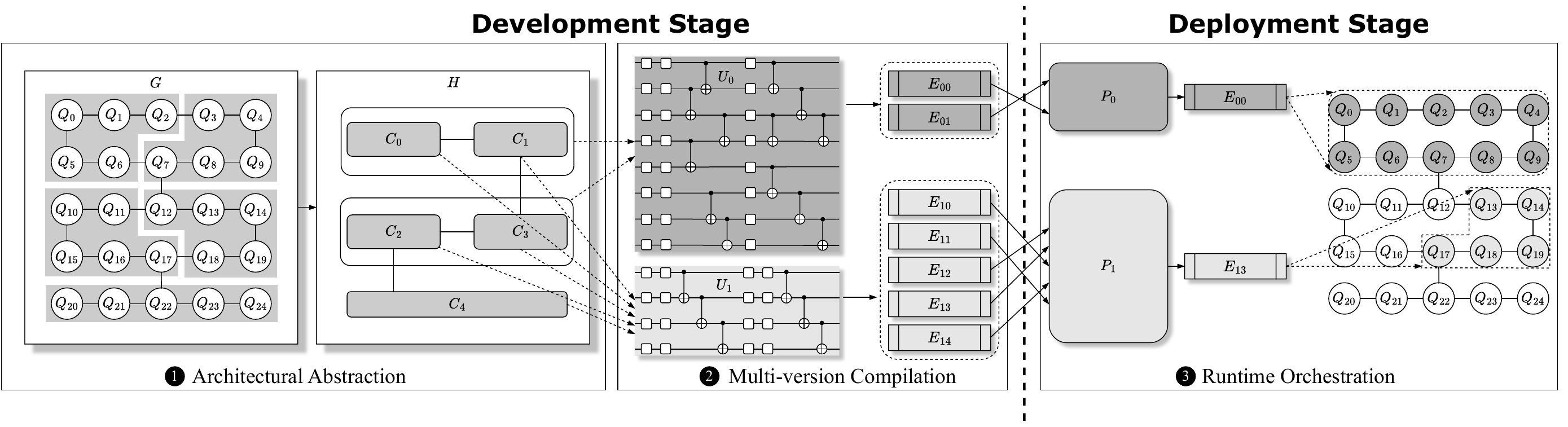}
    \caption{An overview of \name{}. We summarize the notation used in this paper in Table~\ref{tab:symbol_def}.}
    \Description{System overview of FLAMENCO. Offline compilation partitions a target device into compute units and produces multiple physical-region executable versions for each program. At invocation time, the runtime selects one conflict-free executable version per concurrent program while favoring versions with better predicted fidelity.}
    \label{fig:qvm_overview}
\end{figure*}

\begin{table}[b!]
    \centering
    \footnotesize
    \setlength{\tabcolsep}{6pt}
    \resizebox{0.48\textwidth}{!}{%
    \begin{minipage}[t]{0.8\textwidth}
        \centering
        \begin{tabular}{@{}c p{8cm}@{}}
            \toprule
            \textbf{Notation} & \textbf{Definition} \\
            \midrule
            $q$ & Logical qubits in quantum program. \\
            $Q$ & Physical qubits on quantum device. \\
            $U$ & Quantum program ($\S$~\ref{ssec:quantum_basic_and_qc_cloud}). \\
            $G$ & A graph that models a quantum device ($\S$~\ref{ssec:arch_abstraction}). \\
            $C$ & Compute unit ($\S$~\ref{ssec:arch_abstraction}). \\
            $H$ & A graph that represents connections between compute units ($\S$~\ref{ssec:arch_abstraction}). \\
            $m$ & Number of qubits in a compute unit ($\S$~\ref{ssec:arch_abstraction}). \\
            $k$ & Number of logical qubits in a quantum program. \\
            \bottomrule
	\end{tabular}
    \end{minipage}%
    }
    \hfill
    \resizebox{0.48\textwidth}{!}{%
    \begin{minipage}[t]{0.8\textwidth}
        \centering
        \begin{tabular}{@{}c p{8cm}@{}}
            \toprule
            \textbf{Notation} & \textbf{Definition} \\
            \midrule
            $n$ & Number of physical qubits in a quantum device. \\
            $\pi$ & A mapping from $Q$ to $C$ ($\S$~\ref{ssec:arch_abstraction}). \\
            $P$ & Quantum process ($\S$~\ref{ssec:fid_aware_orch}). \\
            $E$ & Quantum executable ($\S$~\ref{ssec:fid_aware_orch}). \\
            $M$ & Number of co-running programs ($\S$~\ref{ssec:fid_aware_orch}). \\
            $K_{\{1,\dots,M\}}$ & Number of executables in each process ($\S$~\ref{ssec:fid_aware_orch}). \\
            $X_{\{1,\dots,M\}}$ & Indices of selected executables ($\S$~\ref{ssec:fid_aware_orch}). \\
            $L_{\{C,P,E\}}$ & List of compute units, processes or executables. \\
            $\# \texttt{ xxx}$ & Number of \texttt{xxx}. \\
            \bottomrule
        \end{tabular}
    \end{minipage}%
    }
    \caption{Notation used in this paper.}
    \label{tab:symbol_def}
\end{table}

\section{\name{}}

In this section, we present the system design. The notation used throughout this paper is summarized in Table~\ref{tab:symbol_def}.

\subsection{Challenges}

\label{sec:design_challenges}

To remove online compilation from the critical execution path, we ask: \tb{is it feasible to enable multiprogramming for pre-compiled quantum executables?}
Unlike queue-level schedulers that decide which jobs should run, our design focuses on a later decision point: once a set of programs has been selected for co-execution, can the runtime choose one pre-compiled executable per program so that the selected executables are conflict-free and high fidelity?
While this practice is common and natural in classical systems, QC imposes two challenges.

\squishlist{}

\item \textbf{Challenge 1: Enabling Portable Execution.}
Without online compilation,
conflicts over the physical qubits used by different programs may inevitably arise, as the specific regions where other programs reside are unknown.
To address this, it is necessary to have portable compilation outcomes, allowing the runtime to resolve resource conflicts by selecting executables mapped to non-overlapping regions on demand.
However, as discussed in $\S$~\ref{ssec:quantum_compilation}, the characteristics of the physical qubits, including their connectivity and error rates, heavily influence the outcome of compilation.
The inherent variability~\cite{smithScalingSuperconductingQuantum2022} present in current hardware exacerbates the dependency on qubit properties, making it impractical to directly execute a program on a different region.

\item \textbf{Challenge 2: Preserving High-Fidelity Execution.}
Previous studies have demonstrated that the fidelity of multi-program execution programs is lower compared to single-program execution~\cite{dasCaseMultiProgrammingQuantum2019}.
The essential reason is that, while a single program can often use its best available region, concurrent execution may force some programs onto suboptimal regions to avoid conflicts.
To mitigate this issue, previous works propose various online co-optimizations for co-running programs~\cite{liuQuCloudNewQubit2021,liuQuCloudHolisticQubit2022,niuHowParallelCircuit2022}.
However, preserving overall fidelity becomes challenging when programs are compiled independently offline.
This challenge stems from the inability of offline compilation to incorporate online co-optimizations for all concurrent programs.
Additionally, the fidelity of program execution greatly depends on the mapped region, but it is difficult to determine the mapped region offline due to the probable conflicts with other programs, which adds an extra layer of complexity for this problem.

\squishend{}

In the subsequent sections, we introduce this design to address these challenges.
The key design principle is to separate development, deployment, and invocation. During development, each structurally stable quantum workload is compiled into a library of physical-region executable versions. During deployment, this executable library is made available as part of the quantum service. During each post-deployment invocation, after an external scheduler has selected a co-execution set, the runtime chooses one version per program. This design makes multi-version compilation a one-time development cost and amortizes it across long-term repeated invocations.

\begin{figure}
    \captionsetup[subfigure]{belowskip=-0.5pt}
    \centering
    \begin{subfigure}{0.22\textwidth}
        \includegraphics[width=0.99\linewidth]{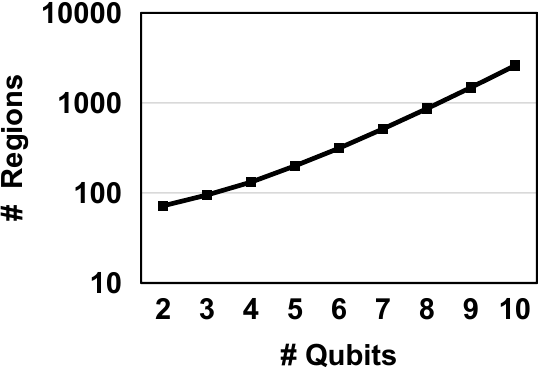}
        \caption{\# Regions vs. \# qubits.}
        \label{fig:num_parts_vs_qubits}
    \end{subfigure}
    \begin{subfigure}{0.7\textwidth}
        \includegraphics[width=0.99\linewidth]{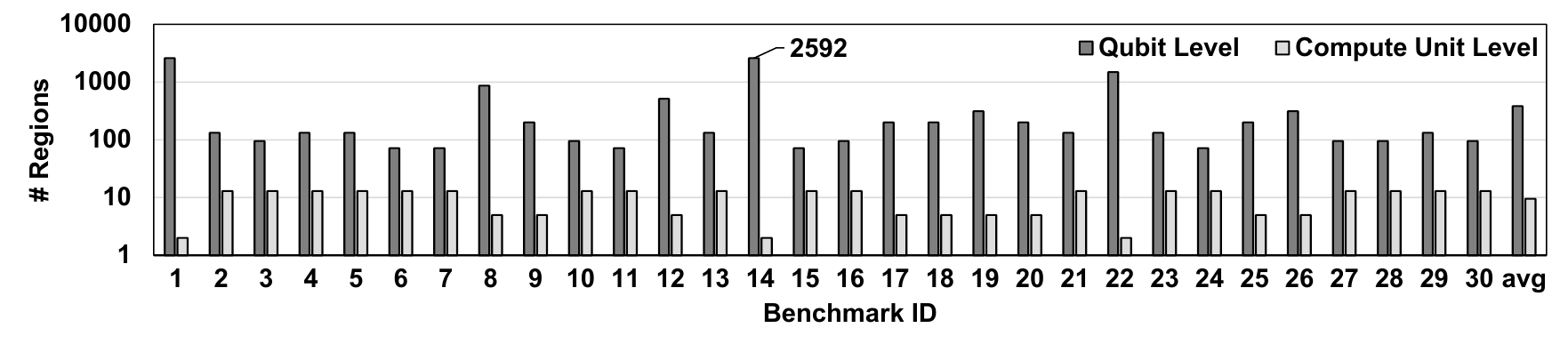}
        \caption{Comparison between qubit level allocation and compute unit level allocation (m=4).}
        \label{fig:num_parts_cmp_qubit_cu}
    \end{subfigure}
    \caption{(a) Qubit level allocation results in exponential growth of \# regions with the increase of \# qubits. (b) Compute unit level allocation significantly reduces \# regions by around two orders of magnitude compared to qubit level allocation scheme.}
    \Description{Two plots compare candidate-region counts. Qubit-level allocation produces exponentially more regions as device size grows, while allocation with four-qubit compute units reduces the number of regions by roughly two orders of magnitude.}
\end{figure}

\subsection{Architectural Abstraction}
\label{ssec:arch_abstraction}

\subsubsection{Design Considerations}

To address \textbf{Challenge 1}, the central concept of our design is straightforward: employing \emph{multi-version compilation},
\ie, compiling a program on diverse qubit regions to produce multiple versions of executables.
This approach allows the concrete executable region of each already-selected program to be determined dynamically at runtime for multi-program execution.
However, it is crucial to strike a balance in the number of versions required.
Excessive versions introduce a huge search space for orchestration at runtime ($\S$~\ref{ssec:fid_aware_orch}), which penalizes the overall system performance.
This issue stems from the inherent characteristics of QC, that is, a qubit serves as the basic unit of both computation and information storage.
To execute a $k$-qubit program on an $n$-qubit device, \emph{every} connected region of $k$ qubits within the device can potentially be employed.
As a consequence, \emph{fine-grained} allocation at qubit-level results in an exceedingly large number of candidate regions.
To quantify this impact, we conduct experiments on the IBM \texttt{FakeBrooklyn} (\texttt{BKLYN}) device model ($\S$~\ref{ssec:system_conf}).
From Figure~\ref{fig:num_parts_vs_qubits}, we observe that \# regions grows exponentially as \# qubits increases.
For example, there exist 2592 possible regions for a 10-qubit quantum program (Figure~\ref{fig:num_parts_cmp_qubit_cu}).

\begin{tcolorbox}[colback=gray!10!white,colframe=gray!10!black,sharp corners,left=0mm,right=0mm,top=0mm,bottom=0mm,title=The Trade-off between Search Space and Device Utilization]
Employing too few versions may also undermine system utilization due to the higher possibility of resource conflicts among co-running programs and can negatively impact fidelity, as indicated by the analysis in $\S$~\ref{sssec:scale_discuss}.
\end{tcolorbox}

\subsubsection{Resource Partition and Allocation}
\label{ssec:partition}

To reduce the vast search space of candidate versions, we propose \emph{coarse-grained} allocation at \emph{compute unit} level.
A quantum device is modeled as a graph, denoted as $G$, where each node represents a physical qubit, each edge denotes a physical connection, or a link, between two qubits, and the edge weight is the calibrated two-qubit operation error rate~\cite{liTacklingQubitMapping2019}.
A compute unit is defined as a connected subgraph within $G$, denoted as $C$.
Initially, the quantum device is divided into multiple compute units, each consisting of $m$ qubits.
When mapping a specific $k$-qubit program, we allocate $\lceil \frac{k}{m} \rceil$ connected compute units for its execution.
Consequently, the number of available regions for compilation is significantly reduced (Figure~\ref{fig:num_parts_cmp_qubit_cu}), from an average of 384.3 to 9.5.

The procedure of generating compute units is detailed in Algorithm~\ref{alg:cu_gen}.
We commence by calculating the \emph{utility} value for each physical qubit, defined as $\frac{\text{\# links}}{\text{Sum of link error}}$~\cite{dasCaseMultiProgrammingQuantum2019}.
Taking the insight that high-utility nodes are typically sparsely distributed~\cite{dasCaseMultiProgrammingQuantum2019,liuQuCloudNewQubit2021}
, multiple high-quality compute units can be generated by iteratively applying breadth-first search (BFS) on $G$.
More specifically,
when unvisited nodes remain, we select the highest utility node as the root and expand a subgraph containing $m$ qubits from root to form a compute unit (lines 4--6).
This expansion is a standard BFS over the remaining graph: starting from the root, it visits unvisited neighbors level by level and stops after collecting \(m\) connected qubits.
Throughout this process, we also record the interrelations among different compute units in $H$ (lines 10--13), essentially a graph with compute units as nodes, where edges in $H$ indicate inter-unit connections.
For larger quantum programs, we then apply the same procedure from Algorithm~\ref{alg:cu_gen} to graph $H$ to form multiple qubit regions, where the utility of a compute unit is calculated by summing the utilities of its constituent qubits.
For example, a 25-qubit device is abstracted into 5 compute units, each with 5 qubits (Figure~\ref{fig:qvm_overview} \circled{1}).
An 8-qubit quantum program $U_0$ requires 2 compute units, then we may allocate $\{C_0,C_1\}$ and $\{C_2,C_3\}$ for this program.
The allocation scheme implies that larger programs may have fewer available regions than smaller ones.
For example, program $U_0$ has two candidate regions, while a smaller program $U_1$ has five (Figure~\ref{fig:qvm_overview} \circled{2}).
The impact of this characteristic is further discussed in $\S$~\ref{sssec:scale_discuss}.
The value of \( m \) determines the number of candidate regions and also affects execution fidelity; we evaluate this impact in $\S$~\ref{ssec:impact_cu_size}.
The scalability and version-library-overhead implications of this compute-unit design are discussed in $\S$~\ref{sec:discussion}.

\begin{figure}[t]
\centering
\resizebox{0.48\textwidth}{!}{%
\begin{minipage}[t]{0.6\textwidth}
\begin{algorithm}[H]
    \footnotesize
    \SetAlgoLined
    \SetKwFunction{getonecu}{get\_one\_cu}
    \SetKwFunction{append}{append}
    \SetKwFunction{getqubits}{get\_qubits}
    \SetKwInOut{KwIn}{Input}
    \SetKwInOut{KwOut}{Output}
    \KwIn{$m$ \tcp{Number of qubits in a compute unit.}}
    \KwIn{$G$ \tcp{A graph that models a quantum device. Each node corresponds to a physical qubit, each edge represents a connection between two qubits. Edge weights represent 2-qubit operation error rates.} }
    \KwOut{$L_C$ \tcp{List of compute units.}}
    \KwOut{$H$ \tcp{A graph that models the connections between compute units.}}

    $\pi$ = $\emptyset$ \tcp{Initialize the mappings between physical qubits and corresponding compute units.}

    Calculate the utility of each node in $G$.

    \tcp{Generate compute units and record the mapping between qubits and compute units.}
    \While{there remain unvisited nodes in $G$}{
        Set the node with the highest utility as root \\
        \emph{$C$} $\gets$ grow a subgraph with $m$ qubits from the root \\
        Remove nodes in $C$ from $G$ and re-calculate utilities \\
        \For{$Q$ in $C$}{
            $\pi$[$Q$] = \emph{$C$} \\
            $L_C$.\append{\emph{$C$}}
        }
    }

    \tcp{Traverse the edges in $G$ and construct $H$.}
    \For{edge in $G$}{
        $Q_0$, $Q_1$ $\gets$ qubits connected by the edge \\
        $C_0$, $C_1$ = $\pi$[$Q_0$], $\pi$[$Q_1$] \\
        $H$[$C_0$].\append{$C_1$}
    }
    \caption{Compute units generation.}
    \label{alg:cu_gen}
\end{algorithm}
\end{minipage}%
}\hfill
\resizebox{0.48\textwidth}{!}{%
\begin{minipage}[t]{0.6\textwidth}
\begin{algorithm}[H]
    \footnotesize
    \SetAlgoLined
    \SetKwFunction{getcus}{get\_comp\_units}
    \SetKwFunction{append}{append}
    \SetKwInOut{KwIn}{Input}
    \SetKwInOut{KwOut}{Output}
    \KwIn{\emph{$L_{P}$} \tcp{List of processes.}}
    \KwOut{\emph{$L_{E}$} \tcp{List of selected executables.}}

    \tcp{Firstly, we rank executables based on their individual costs, this process can be done offline.}
    \For{$P$ in $L_{P}$}{
        Rank executables based on their costs
    }

    $L_{C}$ = $\emptyset$ \tcp{Initialize the selected compute units.}

    \tcp{For each process, traverse executables and greedily choose an executable whose compute units are not in $L_{C}$.}
    \For{$P$ in $L_{P}$}{
        \For{$E$ in $P$}{
            \If{$E$ $\cap$ $L_{C}$ == $\emptyset$}{
                $\{C\}$ $\gets$ obtain the corresponding compute units from $E$ \\
                Add all compute units in $\{C\}$ to $L_{C}$ \\

                $L_{E}$.\append{$E$} \\

                \tcp{We have found an executable for current process, move forward to the next one.}
                break
            }
        }
    }
    \caption{Executable selection.}
    \label{alg:exe_selection}
\end{algorithm}
\end{minipage}%
}
\Description{Two pseudocode algorithms. The first partitions a device graph into fixed-size compute units and constructs the graph connecting those units. The second greedily traverses ranked executable versions and selects one conflict-free version for each process.}
\end{figure}

\subsection{Fidelity Evaluation}

\label{ssec:multi_version_compilation}

To tackle \textbf{Challenge 2}, the system provides a \textbf{general and modular framework} for version characterization.
A key design principle is to \textbf{decouple} version characterization from the runtime selection mechanism, allowing the fidelity evaluation function to be easily customized or replaced.
Specifically, multi-version compilation allows us to characterize each executable offline and rank its potential fidelity before runtime orchestration.
This ranking function is implemented as a modular component, supporting different evaluators tailored to specific hardware backends or accuracy requirements.

\noindent
\textbf{Definition~1.}
\emph{Let $E$ be a quantum executable, which is the compilation outcome on a specific qubit region. Let $\mathcal{L}(E)$ represent the cost of an executable $E$ compiled on compute unit $C$. The cost is expressed as a tuple comprising two metrics:
\begin{equation}
\scriptsize
\mathcal{L}(E) = \left(\frac{d_o}{d_i}, \mathcal{U}(C)\right),
\end{equation}
where $d_i$ and $d_o$ respectively indicate the circuit depth prior to and after compilation.
Additionally, $\mathcal{U}(C)$ denotes the sum of utilities~\cite{dasCaseMultiProgrammingQuantum2019} ($\S$~\ref{ssec:partition}) of all qubits in $C$.
}

\noindent
\tb{The intuition} of using circuit depth stems from the observation that it is an important compiler-derived proxy for noise exposure on limited-connectivity devices ($\S$~\ref{ssec:quantum_compilation}).
In this paper, we employ $\frac{d_o}{d_i}$ as a \textbf{default implementation instance} of the cost function, as it provides an effective balance between ranking accuracy and the stringent low-latency requirements of the runtime path.
However, the architecture itself is metric-agnostic; more sophisticated evaluators incorporating two-qubit gate counts, calibrated error rates, or crosstalk estimates can be integrated as modular plug-ins without altering the core orchestration logic.
In cases where two versions yield identical values of $\frac{d_o}{d_i}$, tie-breaking is made based on $\mathcal{U}(C)$.
The effectiveness of this modular approach and the performance of different evaluators are further evaluated in the experimental section ($\S$~\ref{ssec:impact_fid_orch}).

\subsection{Fidelity-Aware Runtime Orchestration}
\label{ssec:fid_aware_orch}

Given a list of quantum programs awaiting processing, queue-level scheduling decides which programs should run and when. This problem is similar to classical non-preemptive scheduling problems~\cite{krauseTaskschedulingAlgorithmMultiprogramming1973} and has also been studied for MPQC~\cite{dasCaseMultiProgrammingQuantum2019,liuQuCloudNewQubit2021,raviAdaptiveJobResource2021}.
In contrast, our design assumes that multiple quantum programs have already been chosen to co-execute on a single device, and focuses on how to designate the exact executable version for each of them (Figure~\ref{fig:qvm_overview} \circled{3}).
We use the term \emph{orchestration} to distinguish this issue from queuing and scheduling.
To formalize this problem, we introduce the concept of quantum \emph{process} defined as follows.

\noindent
\textbf{Definition 2.}
\emph{Let $P$ be a quantum process, which is a list of executables compiled on multiple regions and ranked based on their costs. A list of processes is denoted as $\{P_1,\dots,P_M\}$, each with $\{K_1,\dots,K_M\}$ executables: $\{\{E_{11},\dots,E_{1K_1}\},\dots,\{E_{M1},\dots,E_{MK_M}\}\}$. For every $i \in [1,M]$, we have $\mathcal{L}(E_{i1}) \leq \mathcal{L}(E_{i2}) \leq \cdots \leq \mathcal{L}(E_{iK_i})$.
}

Based on the definition above, our primary objective is to choose an executable $E_{iX_i}$ from each process $P_i$ such that the selected executable set $\{E_{1X_1},\dots,E_{MX_M}\}$ is conflict-free.
Moreover, we hope to select executables with higher execution fidelity.
Since the executables within a process are estimated and ranked in the offline stage, a smaller executable index $X_i$ represents higher potential fidelity.
Consequently, the sum of indexes should be minimized and the optimization objective is as follows.

\begin{equation}
\scriptsize
    \label{eq:orchestration}
    \begin{aligned}
        \text{Minimize} \quad & \sum_{i=1}^M X_i \\
        \text{Subject to:} \quad & E_{iX_i} \cap E_{jX_j} = \emptyset, \quad \forall i \neq j.
    \end{aligned}
\end{equation}

\noindent
To find the optimal solution, we can enumerate all possible combinations of executables from $M$ processes.
Clearly, we have to evaluate $\prod_{i=1}^M K_i$ combinations, which grows exponentially as $M$ increases.
Consequently, finding the optimal solution using this brute-force strategy introduces significant runtime overhead.
To alleviate this issue, we adopt a lightweight heuristic method (Algorithm~\ref{alg:exe_selection}) to select executables from different processes without resource conflicts and with high predicted fidelities.
Specifically, given a list of processes, we traverse the executables within each process in descending order of predicted fidelity (line 5).
Concurrently, we maintain a set $L_{C}$ to record the compute units allocated by all selected executables.
When finding an executable with no intersection with $L_{C}$, we add this executable to $L_{E}$ and proceed to the next process (lines 7--10).
As a result,
the number of combinations to be evaluated is reduced to $\sum_{i=1}^M K_i$.

Compared to enumerating all executable combinations, the heuristic approach may produce suboptimal solutions. This is an explicit trade-off that keeps the orchestration overhead small enough for the runtime critical path.
Specifically, the first selected executable always ranks first in the process.
However, the executables selected later rank lower in their processes due to the higher possibility of conflicts with the selected compute units.
If no conflict-free executable can be selected for every process, the runtime reports failure for that co-execution set, allowing the upstream scheduler to choose another set.

\emph{Remark.} The traversal order of processes employed in the heuristic orchestration strategy may affect execution fidelity.
We explore this impact later in $\S$~\ref{ssec:scalability}.

\emph{Trade-off.} By targeting a forward-looking service-deployment model, the design removes online co-compilation from the per-invocation critical path, but it also gives up inter-program optimizations that require seeing all co-running programs simultaneously, such as global routing, cross-program SWAP insertion, or joint remapping across program boundaries. This trade-off is beneficial when the same structurally stable workload is invoked repeatedly, because development-time physical-region multi-version compilation can be amortized while runtime orchestration remains lightweight.

\section{Methodology}

\subsection{Benchmarks}

\label{ssec:method_bench}

\begin{table}[t!]
    \centering
    \footnotesize
    \setlength{\tabcolsep}{2pt}
    \resizebox{0.48\textwidth}{!}{%
    \begin{minipage}[t]{0.8\textwidth}
        \centering
        \begin{tabular}{@{}c l p{6cm}@{}}
            \toprule
            \textbf{Index} & \textbf{Name} & \textbf{Description} \\
            \midrule
            1  & \texttt{adder\_n10} & Quantum ripple-carry adder~\cite{cuccaro2004new_quantum_adder} \\
            2  & \texttt{adder\_n4}  &  \\
            3  & \texttt{basis\_change\_n3} & Transform the single-particle basis of a linearly connected electronic structure~\cite{PhysRevLett.120.110501_basis_change} \\
            4  & \texttt{basis\_trotter\_n4} & Implement Trotter steps for molecule LiH at equilibrium geometry~\cite{PhysRevLett.120.110501_basis_change} \\
            5  & \texttt{cat\_state\_n4} & Coherent superposition of two coherent states with opposite phase~\cite{Leibfried2005-or_cat_state} \\
            6  & \texttt{deutsch\_n2} & Deutsch algorithm with 2 qubits for $f(x) = x$~\cite{nielsenQuantumComputationQuantum2010} \\
            7  & \texttt{dnn\_n2} & 3 layer quantum neural network sample~\cite{stein2021qasmbench_dnn} \\
            8  & \texttt{dnn\_n8} &  \\
            9  & \texttt{error\_correctiond3\_n5} & Error correction with distance 3 and 5 qubits~\cite{Michielsen2017-sc_error_correctiond3} \\
            10 & \texttt{fredkin\_n3} & Controlled-swap gate~\cite{Patel2016-lc_fredkin} \\
            11 & \texttt{grover\_n2} & Grover’s algorithm~\cite{gilliamGroverAdaptiveSearch2021} \\
            12 & \texttt{hhl\_n7} & HHL algorithm~\cite{harrowQuantumAlgorithmLinear2009} \\
            13 & \texttt{hs4\_n4} & Hidden subgroup problem~\cite{lomont2004hidden_hs4} \\
            14 & \texttt{ising\_n10} & Ising model simulation~\cite{Cervera_Lierta_2018_ising_model_simulation} \\
            15 & \texttt{iswap\_n2} & An entangling swapping gate~\cite{crossOpenQASM3Broader2022a} \\
            16 & \texttt{linearsolver\_n3} & Solver for a linear equation of one qubit~\cite{Branciard2005-bi_linear_solver} \\
            17 & \texttt{lpn\_n5} & Learning parity with noise~\cite{Pietrzak2012-by_lpn} \\
            \bottomrule
        \end{tabular}
    \end{minipage}%
    }
    \hfill
    \resizebox{0.48\textwidth}{!}{%
    \begin{minipage}[t]{0.8\textwidth}
        \centering
        \begin{tabular}{@{}c l p{6cm}@{}}
            \toprule
            \textbf{Index} & \textbf{Name} & \textbf{Description} \\
            \midrule
            18 & \texttt{pea\_n5} & Phase estimation algorithm~\cite{crossOpenQASM3Broader2022a} \\
            19 & \texttt{qaoa\_n6} & Quantum approximate optimization algorithm~\cite{farhiQuantumApproximateOptimization2014} \\
            20 & \texttt{qec\_en\_n5} & Quantum repetition code encoder~\cite{nielsenQuantumComputationQuantum2010} \\
            21 & \texttt{qft\_n4} & Quantum Fourier transform~\cite{nielsenQuantumComputationQuantum2010} \\
            22 & \texttt{qpe\_n9} & Quantum phase estimation algorithm~\cite{nielsenQuantumComputationQuantum2010} \\
            23 & \texttt{qrng\_n4} & Quantum random number generator~\cite{Tamura_2020_qrng} \\
            24 & \texttt{quantumwalks\_n2} & Quantum walks on graphs with up to 4 nodes~\cite{Venegas_Andraca_2012_quantum_walk} \\
            25 & \texttt{shor\_n5} & Shor’s algorithm~\cite{shorPolynomialTimeAlgorithmsPrime1999} \\
            26 & \texttt{simon\_n6} & Simon’s algorithm \\
            27 & \texttt{teleportation\_n3} & Quantum teleportation~\cite{nielsenQuantumComputationQuantum2010} \\
            28 & \texttt{toffoli\_n3} & Toffoli gate~\cite{nielsenQuantumComputationQuantum2010} \\
            29 & \texttt{variational\_n4} & Variational ansatz for a Jellium Hamiltonian with a linear-swap network~\cite{PhysRevLett.120.110501_basis_change} \\
            30 & \texttt{wstate\_n3} & W-state preparation and assessment~\cite{PhysRevA.65.032108_w_state} \\
            31 & \texttt{multiply\_n13} & Perform 3$\times$5 in a quantum circuit \\
            32 & \texttt{cc\_n12} & Quantum counterfeit coin finding~\cite{iwama2012quantum_counterfeit_coin_problem} \\
            33 & \texttt{cat\_state\_n22} & Coherent superposition of two coherent states with opposite phase~\cite{Leibfried2005-or_cat_state} \\
            34 & \texttt{bv\_n14} & Bernstein-Vazirani algorithm~\cite{bernstein1997quantum} \\
            \bottomrule
        \end{tabular}
    \end{minipage}%
    }
    \caption{List of benchmarks. \texttt{\_nxx} denotes the number of qubits in this benchmark quantum program. Throughout this paper, we use the corresponding index (\texttt{ID}) to represent a benchmark, and $\texttt{ID}_1$\_$\dots$\_$\texttt{ID}_M$ to denote a benchmark group with $M$ programs.}
    \label{tab:benchmarks}
\end{table}

We gather a rich set of benchmarks from QASMBench~\cite{liQASMBenchLowLevelQuantum2023}, as listed in Table~\ref{tab:benchmarks}.
There are 34 benchmark programs with \# qubits in the range $[2,22]$ covering different types of programs, including prototype implementations of algorithms (\texttt{shor\_n5}, \texttt{grover\_n2}, \texttt{bv\_n14}) and subroutines (\texttt{qft\_n4}, \texttt{qec\_en\_n5}, \texttt{multiply\_n13}, \etc) targeting future large-scale fault-tolerant machines, as well as variational programs that are potentially practically useful in the NISQ era (\texttt{qaoa\_n6}, \texttt{dnn\_n8}, \etc).
We stop at 22 qubits because our fidelity metric compares output distributions directly, making controlled noisy-simulation validation increasingly expensive at still larger scales.
To evaluate the performance of MPQC when co-executing $M$ programs, we select representative benchmark groups from this benchmark set for each evaluation setting.

\subsection{Implementations}
\label{ssec:implementations}

In this paper, we develop and assess five distinct execution models for running multiple quantum programs:

\squishlist{}
    \item \textbf{Single-program Reference:} Each program is compiled and executed sequentially on the target device, and we report the average fidelity. This reference estimates the fidelity loss introduced by multi-program execution.
    \item \textbf{Baseline:} In line with the state-of-the-art MPQC~\cite{liuQuCloudNewQubit2021}, we employ a community detection assistant partitioning algorithm for the allocation of qubit regions. When multiple programs have adjacent regions, they are amalgamated into a single program and compiled collectively, enabling inter-program optimizations.
    \item \textbf{Vanilla:} This model uses architectural abstraction and multi-version compilation. However, it does not estimate the potential fidelity of executables, instead opting to randomly select conflict-free executables at runtime. The vanilla model acts as a point of comparison to showcase the effectiveness of our design in mitigating fidelity loss.
    \item \textbf{\name:} This model is the complete embodiment of our proposed design. It differs from the vanilla model by incorporating the lightweight cost function during compilation and executing fidelity-aware orchestration at runtime. As a result, it helps reduce fidelity loss compared to the vanilla implementation.
    \item \textbf{\name{}-Pro:} This model keeps the same architectural abstraction, multi-version compilation, and runtime orchestration workflow as \name{}, but replaces the default lightweight cost function with a richer physically informed lexicographic ranking rule. Specifically, candidate executables are ranked by lower estimated physical error \ie, lower compiled gate-error sum plus readout-error sum, then by lower compiled SWAP count, lower compiled two-qubit-gate count, lower compiled depth, lower compiled-size ratio, and finally higher compute-unit quality. This model is used to evaluate whether the system's post-compilation ranking rule can be upgraded without changing the overall orchestration interface.
\squishend{}

To maintain a fair comparison, the default compilation method in Qiskit is adopted for all implementations.
All experiments in this paper use Qiskit 0.39.0 for circuit compilation and simulation.
The system serves as a compilation manager that can integrate any user-defined compilation methods.
In addition, crosstalk-aware strategies~\cite{niuHowParallelCircuit2022} are omitted since they are complementary to both the baseline and the full system.
The impact of crosstalk is discussed in $\S$~\ref{ssec:crosstalk}.

\subsection{Metrics}
\label{ssec:metrics}

We evaluate the performance of the system in terms of latency, fidelity, and utilization based on the following metrics.

\squishlist{}
    \item \textbf{Latency:} As emphasized in $\S$~\ref{ssec:problem_and_motivation}, we focus on the runtime of programs \emph{after} they are scheduled to be co-executed.
    For this latency metric, we isolate the critical-path software overhead before QPU execution.
    We define \emph{Post-Scheduling Non-QPU Speedup} as \(T_{\texttt{compile}}^{\texttt{baseline}} / T_{\texttt{orch}}^{\text{\name}}\), where the numerator is the baseline's online compilation time and the denominator is \name{}'s runtime orchestration time after the co-execution set has been selected. This metric measures the reduction in non-QPU critical-path time, rather than an acceleration of offline compilation.
    CACTUS uses 20 ns (40 ns) for single- (two-) qubit gates and 300 ns for measurements.

    \item \textbf{Fidelity:}
    Our goal is to quantify the similarity between two probability distributions produced by an ideal reference and a noisy execution. For interpretability, the metric should range in $[0,1]$ and assign higher values to greater similarity.
    We adopt an exponential KL-based similarity metric between an ideal distribution $P$ and an observed distribution $Q$ over the outcome set $S$: $\mathcal{F}(P,Q)=\exp(-D_{\mathrm{KL}}(P\|Q))$, where $D_{\mathrm{KL}}(P\|Q)=\sum_{s\in S} P(s)\log\frac{P(s)}{Q(s)}$.
    We apply a small smoothing constant before normalization so that missing support in $Q$ is penalized but does not make the metric undefined. The value lies in $[0,1]$, where $1$ indicates identical distributions and values closer to $0$ indicate lower similarity.
    \item \textbf{Utilization:}
    The effectiveness of utilization improvements by MPQC designs has been thoroughly investigated in previous works~\cite{dasCaseMultiProgrammingQuantum2019,liuQuCloudNewQubit2021,liuQuCloudHolisticQubit2022,niuHowParallelCircuit2022}.
    Therefore, we skip this comparison with the single-program reference and focus on the comparison between the proposed system and the baseline.
    The utilization is evaluated based on the conflict-free orchestration ratio (COR), defined as the fraction of evaluated benchmark groups for which the runtime finds one conflict-free executable version for every program in the selected co-execution set: $\mathrm{COR} = \frac{\text{\# groups with conflict-free executable orchestration}}{\text{\# evaluated benchmark groups}}$.

\squishend{}

\begin{figure*}[t!]
    \centering
    \includegraphics[width=\textwidth]{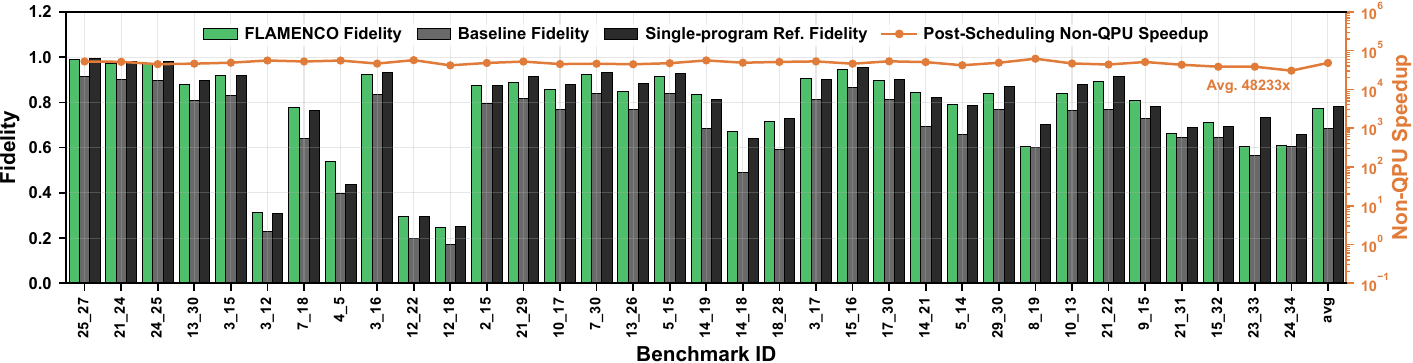}
    \caption{Fidelity comparison among \name{}, baseline, and the single-program reference on \texttt{BKLYN}, together with \name{}'s Post-Scheduling Non-QPU Speedup over the baseline on a log-scaled right y-axis.}
    \Description{Grouped results on the BKLYN backend comparing fidelity for FLAMENCO, the online multiprogramming baseline, and the single-program reference, with FLAMENCO's post-scheduling non-QPU speedup shown on a logarithmic secondary axis.}
    \label{fig:fid_bklyn_revision}
\end{figure*}

\begin{figure*}[t!]
    \centering
    \includegraphics[width=\textwidth]{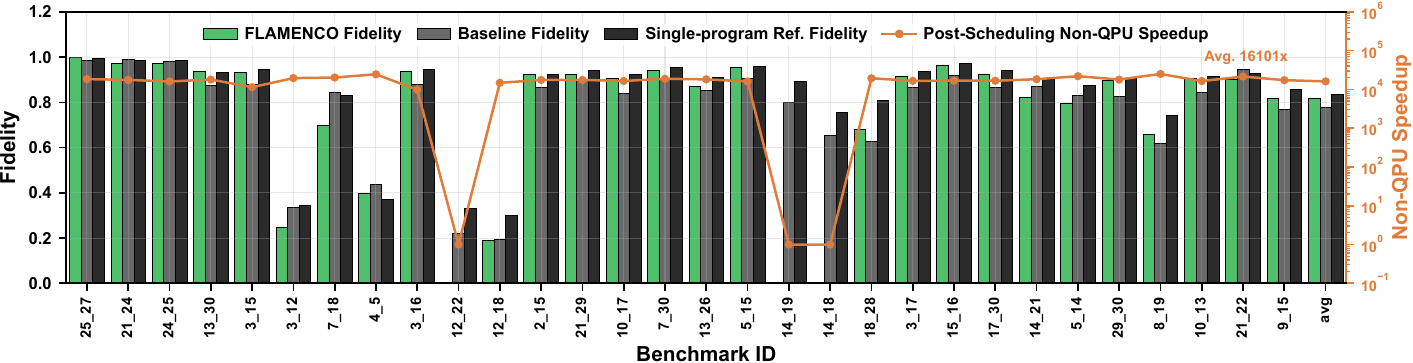}
    \caption{Fidelity comparison among \name{}, baseline, and the single-program reference on \texttt{CAI}, together with \name{}'s Post-Scheduling Non-QPU Speedup over the baseline on a log-scaled right y-axis.}
    \Description{Grouped results on the CAI backend comparing fidelity for FLAMENCO, the online multiprogramming baseline, and the single-program reference, with FLAMENCO's post-scheduling non-QPU speedup shown on a logarithmic secondary axis.}
    \label{fig:fid_cai_revision}
\end{figure*}

\section{Evaluation}

\subsection{\texorpdfstring{Experimental Setup}{Experimental Setup}}
\label{ssec:system_conf}

We perform experiments on both noisy simulators and real quantum devices.
For simulations, we use Qiskit 0.39.0~\cite{Qiskit} with noise models from two IBM quantum devices: \texttt{FakeCairo} (\texttt{CAI}) with 27 qubits and \texttt{FakeBrooklyn} (\texttt{BKLYN}) with 65 qubits.
These simulations are carried out on a Linux server with 2 10-core Intel(R) Xeon(R) CPU E5-2630 v4 CPUs and 128 GB memory.
For real-device validation, we use the publicly available IBM device \texttt{ibm\_osaka} with 127 qubits.
For fidelity evaluation, each program is executed with $2^{20}$ trials to obtain high-resolution estimates of the output probability distributions, and the compute unit size is set to 4 for the full system and vanilla.
As defined in $\S$~\ref{ssec:metrics}, our latency evaluation focuses on post-scheduling runtime overhead, i.e., the critical-path software cost after the co-execution set has already been selected.

\subsection{\texorpdfstring{Main Results}{Main Results}}

\subsubsection{\texorpdfstring{Latency and Fidelity against Online MPQC}{Latency and Fidelity against Online MPQC}}

Figures~\ref{fig:fid_bklyn_revision} and~\ref{fig:fid_cai_revision} show that the proposed system removes online compilation from the post-scheduling critical path and therefore delivers the main benefit of this design: substantially reduced critical-path overhead once the co-execution set is determined.
Across the evaluated workloads, the average Post-Scheduling Non-QPU Speedup is $16101\times$ on \texttt{CAI} and $48233\times$ on \texttt{BKLYN}. Figure~\ref{fig:compilation_vs_duration} further shows that this reduction in online compilation translates to more than 5$\times$ lower post-scheduling runtime.
The same figures also show that this latency reduction does not come with systematic fidelity degradation. Relative to the baseline, the system achieves average fidelities of 0.775 versus 0.687 on \texttt{BKLYN} and 0.818 versus 0.778 on \texttt{CAI}, while the single-program reference provides additional context for the gap between multi-program and isolated execution.

\subsubsection{\texorpdfstring{Fidelity-Aware Orchestration}{Fidelity-Aware Orchestration}}
\label{ssec:impact_fid_orch}

\begin{figure*}[t!]
    \centering
    \includegraphics[width=\textwidth]{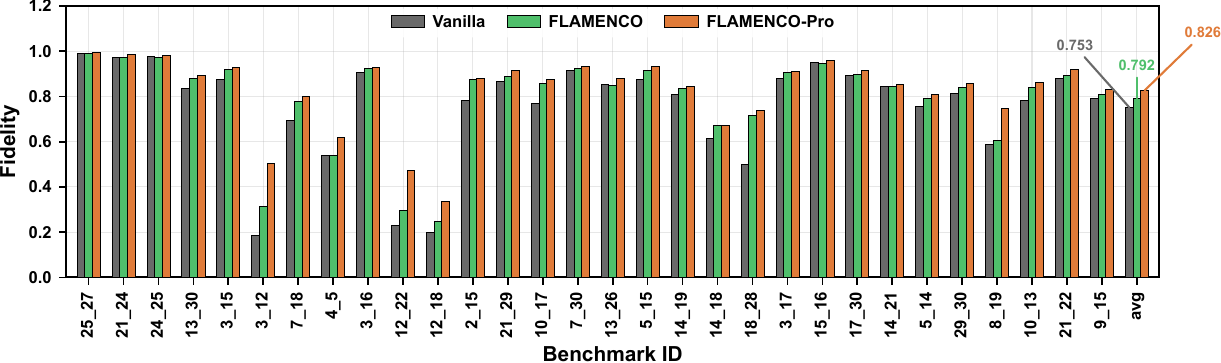}
    \caption{Fidelity comparison among \name{}, vanilla, and \name{}-Pro on \texttt{BKLYN}.}
    \Description{Fidelity results on BKLYN compare FLAMENCO with its vanilla variant and FLAMENCO-Pro across benchmark groups, showing the effect of fidelity-aware executable ranking and selection.}
    \label{fig:fid_comp_vanilla}
\end{figure*}

\begin{figure*}[t!]
    \begin{minipage}[t]{0.49\textwidth}
        \centering
        \includegraphics[width=\linewidth]{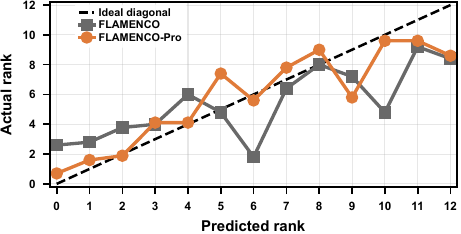}
        \captionof{figure}{Predicted vs. actual fidelity rankings of candidate executable versions on \texttt{BKLYN}. Values are averaged across the sampled 10 benchmarks; lower rank is better.}
        \Description{Scatter plot comparing predicted and measured ranks of candidate executable versions for ten BKLYN benchmarks. Points nearer the diagonal indicate more accurate ranking, and lower ranks indicate better candidates.}
        \label{fig:fid_ranks}
    \end{minipage}\hfill
    \begin{minipage}[t]{0.49\textwidth}
        \centering
        \includegraphics[width=\linewidth]{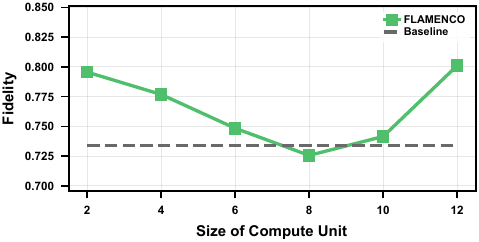}
        \captionof{figure}{Impact of the compute unit size on fidelity.}
        \Description{Line plot of average fidelity on the CAI backend as compute-unit size ranges from two to twelve qubits. Fidelity decreases through size eight and then rises for larger units.}
        \label{fig:impact_cu_size}
    \end{minipage}
\end{figure*}

After establishing in $\S$~5.2.1 that the primary benefit is lower post-scheduling runtime without a systematic fidelity penalty, we next isolate where the remaining fidelity benefit comes from. We first compare the full system with its vanilla version, which keeps the same compute-unit abstraction and multi-version executable library but removes fidelity-aware runtime selection. Figure~\ref{fig:fid_comp_vanilla} shows that multi-version compilation alone is not sufficient: the candidate library only creates a choice space, and the fidelity benefit appears only when the runtime uses that choice space to prefer better executable versions.
Figure~\ref{fig:fid_ranks} evaluates the post-compilation ranking rule on 10 sampled \texttt{BKLYN} benchmarks with 4096 shots per candidate execution. For each benchmark, we rank candidate executable versions by measured fidelity and compare this actual rank with the ranks predicted by \name{} and \name{}-Pro; smaller ranks are better, and values closer to the diagonal indicate better prediction. The default ranking rule is coarse but already places high-quality candidates near the top often enough to improve runtime outcomes. With the richer \name{}-Pro ranking rule from $\S$~\ref{ssec:implementations}, the Top-1 hit rate increases from 0.40 to 0.70, and the Top-3 hit rate increases from 0.90 to 1.00, showing that the ranking component can benefit from richer post-compilation ranking rules without changing the orchestration interface.

\subsubsection{\texorpdfstring{Sensitivity to Compute Unit Size}{Sensitivity to Compute Unit Size}}
\label{ssec:impact_cu_size}

We also examine whether the main fidelity trend depends strongly on a single compute-unit configuration.
Experiments are conducted on the \texttt{CAI} backend using compute unit sizes ranging from 2 to 12 qubits, and we report the average fidelity across 30 benchmark groups.
As shown in Figure~\ref{fig:impact_cu_size}, fidelity first declines as the unit size increases from 2 to 8 and then improves from 8 to 12.
This trend reflects a trade-off: smaller units expose more candidate regions for version generation, while larger units can give small programs a larger local search space once the unit size exceeds most program sizes.
At the same time, larger unit sizes reduce parallelism because they increase the chance of region conflicts.

\subsection{\texorpdfstring{Integration with HyperQ and QOS}{Integration with HyperQ and QOS}}

We next evaluate the proposed system as a post-scheduling enhancement beneath recent scheduler-coupled systems rather than as a drop-in scheduler replacement.
Accordingly, we compare HyperQ with HyperQ+\name{} and QOS with QOS+\name{}, while keeping each base system's original co-execution-set selection policy unchanged.
The only difference is whether the selected co-execution set is realized through a single committed executable/layout or through runtime selection from a multi-version executable library.
We therefore evaluate fixed co-execution workloads rather than queue waiting time or end-to-end scheduling behavior.
We report workload-level COR and common-success fidelity, where fidelity is averaged only over workloads completed by both variants of the same base system.

\begin{figure}[t!]
    \begin{subfigure}[t]{0.60\textwidth}
        \centering
        \includegraphics[width=\linewidth]{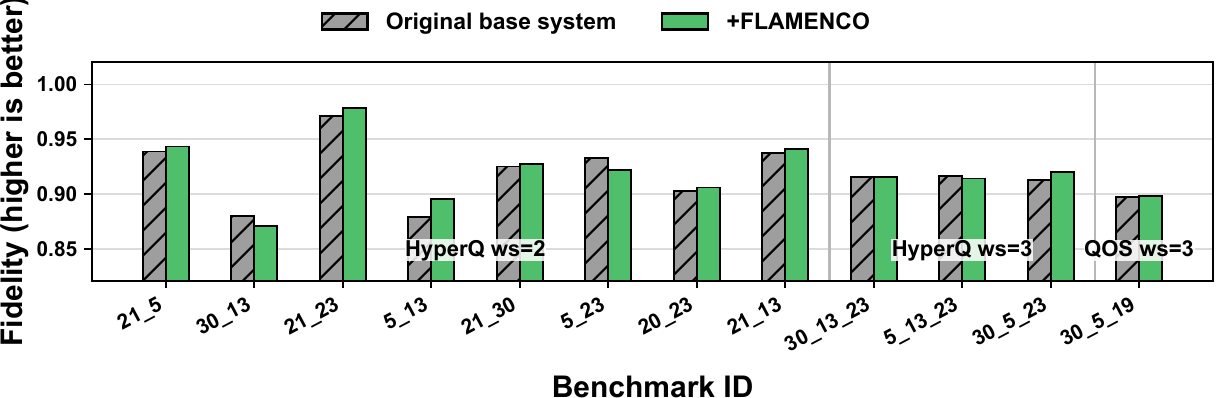}
        \caption{Common-success fidelity on \texttt{BKLYN}.}
        \label{fig:host_integration_fidelity}
    \end{subfigure}
    \hfill
    \begin{subfigure}[t]{0.37\textwidth}
        \centering
        \includegraphics[width=\linewidth]{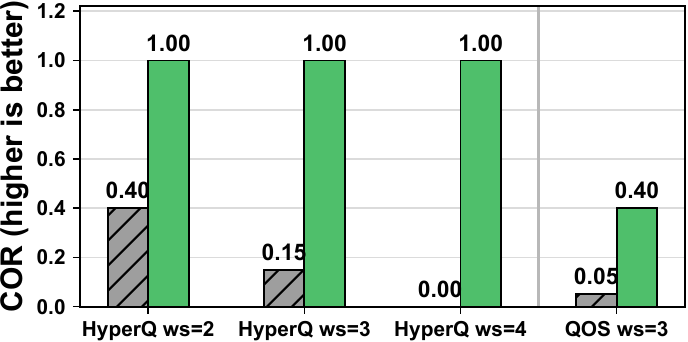}
        \caption{Workload-level COR on \texttt{BKLYN}.}
        \label{fig:host_integration_cor}
    \end{subfigure}
    \caption{Integration with HyperQ and QOS. \name{}-style multi-version executable libraries are added to the same base system without changing the base system's own co-execution-set selection policy. Here, \texttt{ws} denotes workload size, i.e., the number of concurrent programs in each co-execution workload.}
    \Description{Two panels compare HyperQ and QOS with and without FLAMENCO-style multi-version executable libraries. The first reports common-success fidelity and the second reports conflict-free orchestration ratio for workload sizes labeled ws, where ws is the number of concurrent programs.}
    \label{fig:host_integration}
\end{figure}

\paragraph{\texorpdfstring{Results.}{Results.}}\label{sssec:host_results}
The primary gain appears in COR rather than in common-success fidelity.
As shown in Figure~\ref{fig:host_integration}, HyperQ+\name{} raises COR from 0.183 to 1.000 after aggregating workloads of sizes 2, 3, and 4, while QOS+\name{} raises COR from 0.050 to 0.400 on 3-program workloads.
The HyperQ trend is especially clear as the number of concurrent programs increases: COR improves from 0.400 to 1.000 for \texttt{ws}=2, from 0.150 to 1.000 for \texttt{ws}=3, and from 0.000 to 1.000 for \texttt{ws}=4.
By contrast, common-success fidelity remains comparable in both comparisons.
Under the common-success criterion, HyperQ changes from 0.919 to 0.921 across the 11 workloads completed by both variants, while QOS changes from 0.898 to 0.898 on the single workload completed by both variants.

\paragraph{\texorpdfstring{Interpretation.}{Interpretation.}}\label{sssec:host_discuss}
These gains arise because multi-version executable libraries enlarge the local choice space after the co-execution set has already been selected.
The enhanced system can therefore recover conflict-free realizations without changing the base system's upstream scheduling policy.
The effect is stronger for HyperQ because its qVM-level placement benefits more directly from richer mapped executable choices, whereas QOS remains partially constrained by its upstream compatibility-driven bundling decisions.
Overall, these results validate \name{} as a modular backend enhancement that complements higher-level queue management. It improves the quality of an already selected co-execution set rather than replacing queue- or batch-level scheduling.

\begin{figure}[t!]
    \captionsetup[subfigure]{aboveskip=-1pt,belowskip=-1pt}
    \begin{subfigure}{0.49\textwidth}
        \centering
        \includegraphics[width=\linewidth]{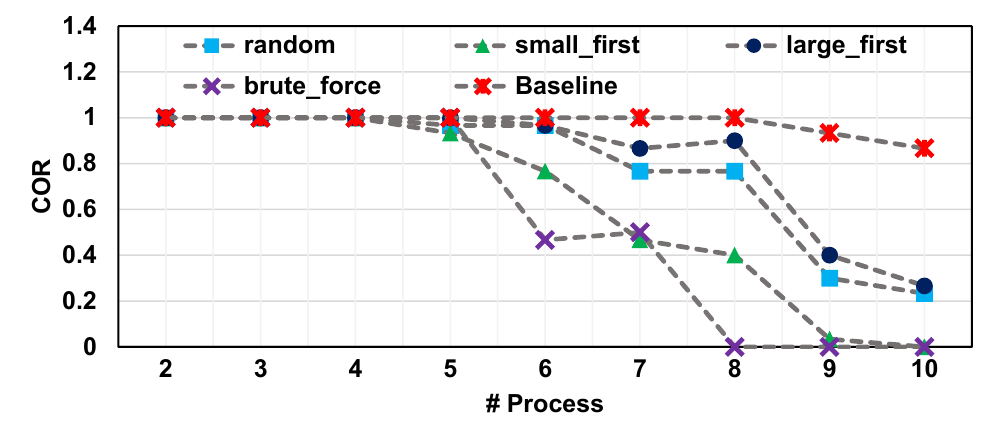}
        \caption{Impact of process count on utilization.}
        \label{fig:utilization}
    \end{subfigure}
    \begin{subfigure}{0.49\textwidth}
        \centering
        \includegraphics[width=\linewidth]{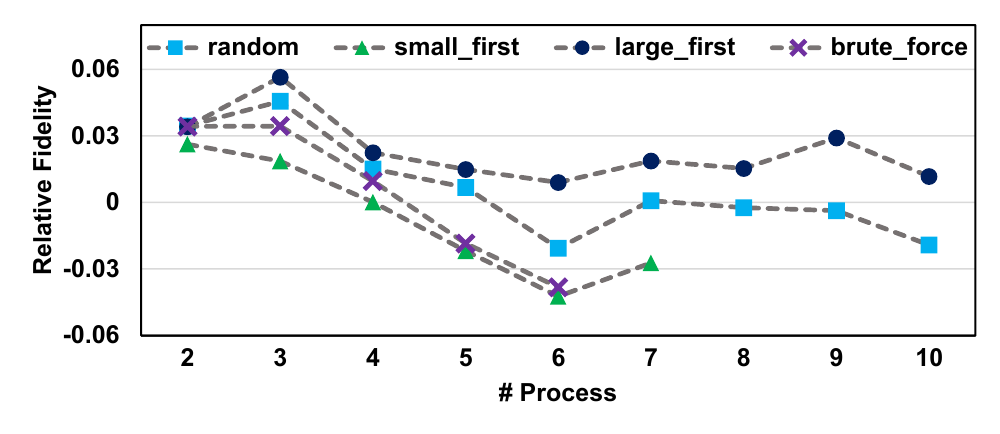}
        \caption{Relative fidelity over baseline.}
        \label{fig:scalability}
    \end{subfigure}
    \caption{Scalability of utilization and relative fidelity as the number of co-running processes increases.}
    \Description{Two scalability plots vary the number of concurrent processes from two to ten. The left reports conflict-free orchestration ratio for several traversal policies and the online baseline; the right reports each policy's relative fidelity over the baseline.}
    \label{fig:scalability_utilization}
\end{figure}

\subsection{\texorpdfstring{Scaling to More Programs}{Scaling to More Programs}}
\label{ssec:scalability}

We use larger co-execution groups to stress-test runtime orchestration.
For each $M \in [2,10]$, we generate 30 benchmark groups using the method in $\S$~\ref{ssec:method_bench} and evaluate four traversal policies.
\texttt{random} traverses processes in random order and reports averages over multiple seeds; \texttt{small\_first} and \texttt{large\_first} traverse processes in ascending and descending circuit-width order, respectively; and \texttt{brute\_force} exhaustively enumerates executable combinations to find the conflict-free combination with the lowest index sum, subject to a 10-second timeout.
To compare utilization with the online-MPQC baseline, we report COR for these 30 groups.
Figure~\ref{fig:utilization} compares COR against the online-MPQC baseline, and Figure~\ref{fig:scalability} reports relative fidelity over the same baseline, defined as $(\mathcal{F}_{\text{policy}}-\mathcal{F}_{\text{baseline}})/\mathcal{F}_{\text{baseline}}$. Values above zero indicate higher fidelity than the baseline, while values below zero indicate lower fidelity. Missing values indicate that the corresponding policy either cannot find conflict-free executables or exceeds the 10-second orchestration timeout.

\paragraph{\texorpdfstring{Results.}{Results.}}\label{sssec:impact_utilization_fidelity}
Figure~\ref{fig:utilization} shows that both the proposed system and the baseline can lose parallel-execution opportunities as the number of co-running programs increases.
Among the evaluated policies, \texttt{large\_first} achieves the highest COR.
For example, its conflict-free orchestration ratio is 26.7\% when $M=10$, whereas the baseline reaches 86.7\%.
Figure~\ref{fig:scalability} shows the corresponding relative-fidelity trends.
While the relative fidelity of the other three strategies gradually decreases and can fall below the baseline, \texttt{large\_first} remains the strongest heuristic among the evaluated policies.
Although \texttt{brute\_force} is exhaustive over executable combinations, it often fails at larger process counts because the exponential search space prevents it from finishing within the timeout; these timeout cases are counted as orchestration failures and lower its COR. Even when it returns a feasible combination, its minimum-index-sum objective can still disagree with true fidelity when the offline ranking is imperfect.

\paragraph{\texorpdfstring{Interpretation.}{Interpretation.}}\label{sssec:scale_discuss}
The main reason is that larger programs have fewer feasible executable versions, as shown in Figure~\ref{fig:num_parts_cmp_qubit_cu}.
Prioritizing smaller programs can therefore leave later large programs with too few conflict-free or high-quality choices, which lowers COR and can also hurt fidelity.
By placing larger programs earlier, \texttt{large\_first} mitigates this version-sparsity effect and therefore gives the most stable scaling behavior among the evaluated traversal policies.
The \(M=10\) case also illustrates the high-utilization trade-off: compute-unit separation can reduce direct inter-program coupling, but reduces packing flexibility and COR relative to the baseline.

\subsection{\texorpdfstring{Hardware Effects}{Hardware Effects}}

\subsubsection{\texorpdfstring{Real-Device Validation}{Real-Device Validation}}
\label{ssec:perf_real_machine}

\begin{figure}[t!]
    \centering
    \begin{minipage}[t]{0.48\textwidth}
        \centering
        \includegraphics[width=\linewidth]{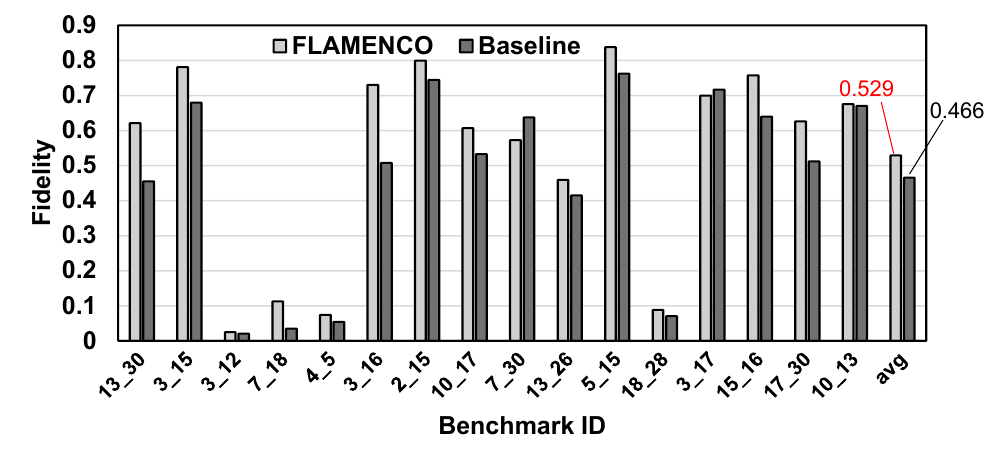}
        \caption{Fidelity comparison between the proposed system and the baseline using \texttt{ibm\_osaka}.}
        \Description{Bar chart comparing FLAMENCO and the online baseline by fidelity for benchmark groups executed on the IBM Osaka quantum device.}
        \label{fig:real_machine}
    \end{minipage}
    \hfill
    \begin{minipage}[t]{0.46\textwidth}
        \centering
        \includegraphics[width=\linewidth]{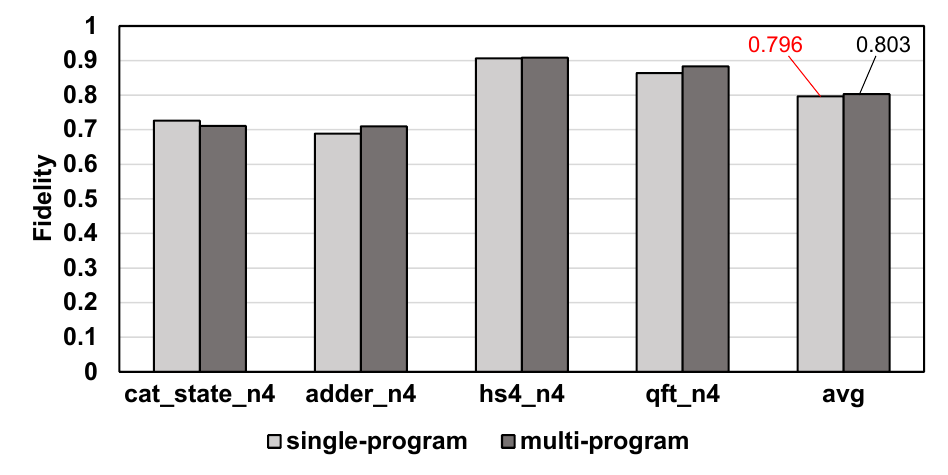}
        \caption{Fidelity comparison between single- and multi-program execution on \texttt{ibm\_osaka}.}
        \Description{Bar chart comparing the fidelity of a program executed alone and concurrently with another program on the same mapped region of the IBM Osaka quantum device.}
        \label{fig:fid_comp_crosstalk_real_machine}
    \end{minipage}
\end{figure}

We include real-device validation to confirm that trends observed in noisy simulation are robust and translate to physical hardware.\footnote{Compared with the simulation study, fewer benchmarks are evaluated on the real device because dynamic programs~\cite{crossOpenQASM3Broader2022a} are not supported and deeply compiled circuits can return invalid results.}
Figure~\ref{fig:real_machine} shows that the real-device results on IBM \texttt{ibm\_osaka} remain consistent with the simulator-based trend, with the system achieving an average fidelity improvement of 13.5\% over the baseline in this experiment.
At the same time, it does not necessarily outperform the baseline on every benchmark, because it gives up online inter-program co-optimization opportunities that require joint compilation of all co-running programs.
We also examine inter-program crosstalk on the same device by comparing single-program and multi-program execution under the same mapped region for $P_0$.
As shown in Figure~\ref{fig:fid_comp_crosstalk_real_machine}, the two fidelities remain at a similar level in this setup, suggesting that the region separation already avoids severe inter-program crosstalk in the evaluated case. Further analysis of crosstalk behavior is provided in $\S$~\ref{ssec:crosstalk}.

\subsubsection{\texorpdfstring{Device Variation}{Device Variation}}
\label{ssec:tolerance_var}

\begin{figure}[h!]
    \captionsetup[subfigure]{aboveskip=2pt,belowskip=-2pt}
    \begin{subfigure}[t]{0.38\textwidth}
        \centering
        \includegraphics[width=\linewidth]{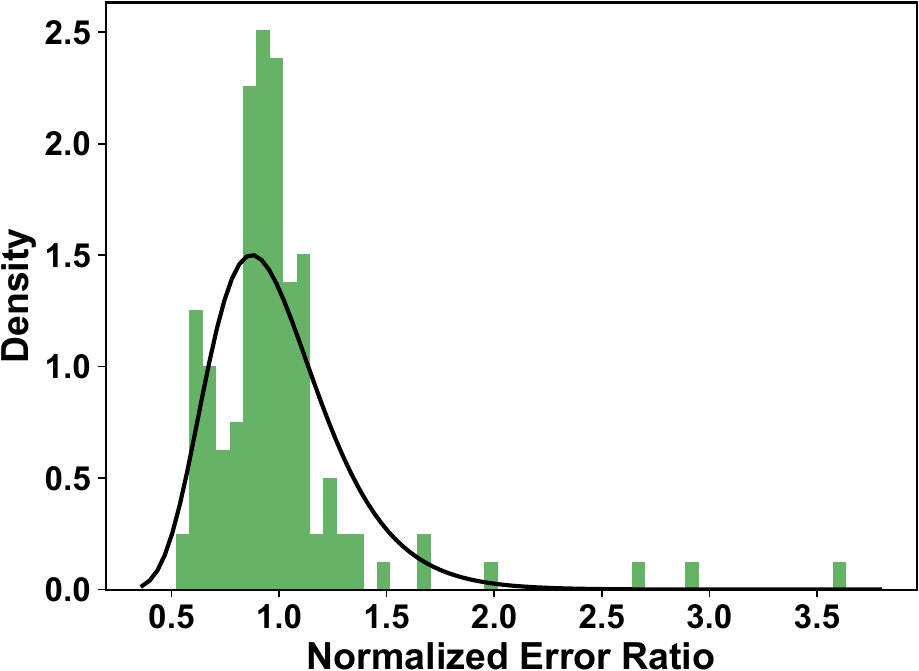}
        \caption{}
        \label{fig:error_ratio_variation}
    \end{subfigure}
    \begin{subfigure}[t]{0.61\textwidth}
        \centering
        \includegraphics[width=\linewidth]{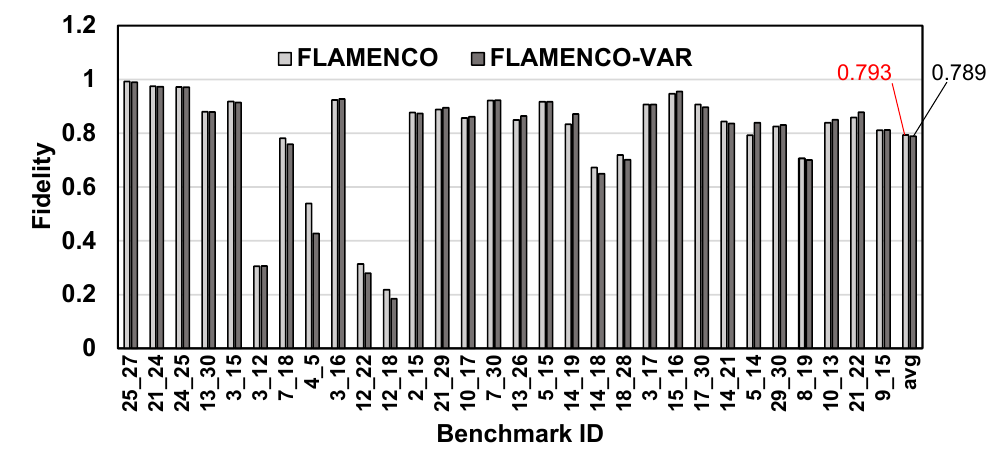}
        \caption{}
        \label{fig:fid_comp_var}
    \end{subfigure}
    \caption{Device variation analysis: (a) probability distribution of normalized two-qubit gate error ratios. The black curve represents the fitted log-normal distribution; (b) fidelity comparison between the standard configuration and \name{}-VAR, demonstrating resilience to device variation.}
    \Description{Two panels analyze device variation. The first shows the distribution of normalized two-qubit gate-error ratios and a fitted log-normal curve. The second compares fidelity under the standard configuration and FLAMENCO-VAR, which uses perturbed offline calibration data.}
    \label{fig:device_variation_analysis}
\end{figure}

Current quantum devices exhibit time-varying characteristics~\cite{muraliFullstackRealsystemQuantum2019}.
To evaluate whether such variation invalidates precompiled executable libraries, we monitor the calibrated two-qubit gate error ratios reported by the IBM \texttt{ibm\_osaka} backend over a continuous ten-day period, normalize them to their overall averages, and fit the resulting distribution with a log-normal model.
We then use the original error ratios for the online stage and perturbed ratios for the offline stage, denoting this setup as \name{}-VAR.
Figure~\ref{fig:fid_comp_var} shows that \name{}-VAR maintains fidelity performance comparable to the standard configuration.
This robustness arises because post-compilation circuit depth depends more strongly on static device topology than on moderate fluctuations in gate error values, and because the relative differences among candidate regions remain largely stable.

\subsubsection{\texorpdfstring{Crosstalk}{Crosstalk}}
\label{ssec:crosstalk}

\begin{figure}[t!]
    \centering
    \includegraphics[width=0.56\linewidth]{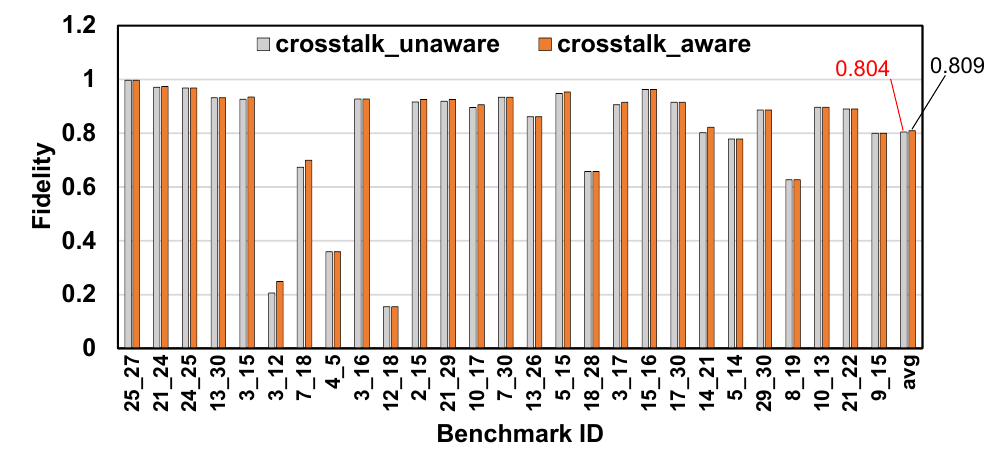}
    \caption{Fidelity comparison between the proposed system with and without crosstalk-aware strategies.}
    \Description{Bar chart comparing fidelity for FLAMENCO with and without a crosstalk-aware executable-selection strategy across benchmark groups.}
    \label{fig:fid_comp_crosstalk_sim}
\end{figure}

Inter-program crosstalk is another hardware effect that can reduce MPQC fidelity~\cite{sarovarDetectingCrosstalkErrors2020}.
Our goal in this experiment is to examine how sensitive the system is to such effects.
To do so, we add a crosstalk-aware strategy on top of the default flow: if a candidate executable places two-qubit operations on high-crosstalk neighboring links, we skip that version and continue the search for another conflict-free one.
Since direct crosstalk characterization on real devices is costly, we use an experimentally validated methodology~\cite{ash-sakiExperimentalCharacterizationModeling2020} to simulate conditional errors and randomly generate crosstalk amplification ratios.
Figure~\ref{fig:fid_comp_crosstalk_sim} shows that this strategy improves fidelity for some benchmarks. This pattern indicates that the strategy itself is effective and that residual inter-program crosstalk risk can still remain in some workloads.
For many other benchmarks, however, the results remain nearly unchanged. This limited incremental effect suggests that the coarse-grained region partitioning already avoids many risky inter-program couplings, making the overall system less sensitive to crosstalk in those cases.
The evaluated programs are also relatively small for the simulated devices, so both \name{} and the baseline often have enough physical space to keep programs apart and start from a low-crosstalk operating point. More tightly packed co-execution sets would increase crosstalk pressure for multiprogramming in general, making the utilization--isolation trade-off more pronounced.

\section{Related Work}
\label{sec:related_work}

Several MPQC designs have been proposed and explored~\cite{dasCaseMultiProgrammingQuantum2019,liuQuCloudNewQubit2021,liuQuCloudHolisticQubit2022,niuHowParallelCircuit2022}.
Das \etal{} introduce MPQC with reliable qubit allocation and delayed instruction scheduling~\cite{dasCaseMultiProgrammingQuantum2019}.
Liu \etal{} use community-detection-based mapping and inter-program SWAP insertion to improve fidelity~\cite{liuQuCloudNewQubit2021}, later extending these techniques to 3D topologies~\cite{liuQuCloudHolisticQubit2022}.
Niu \etal{} present crosstalk-aware parallel execution for applications including variational quantum eigensolver and zero-noise extrapolation~\cite{niuHowParallelCircuit2022}.
These systems optimize within a job-execution lifecycle: after the co-running set is known, they allocate regions and jointly compile or optimize its programs. This model enables inter-program co-optimization but requires the co-running set at compilation time and places compilation on the critical path.

HyperQ, QOS and \name{} differ in execution lifecycle and system layer. HyperQ~\cite{taoQuantumVirtualMachines2025} and QOS~\cite{giortamis2025qosquantumoperating} target the current quantum-cloud job-submission model, in which general jobs are queued on shared QPUs. HyperQ compiles programs independently to quantum virtual machine (qVM) targets, then schedules qVM executables in space and time, places them on compatible regions, and translates them into aggregated machine-sized circuits~\cite{taoQuantumVirtualMachines2025}. QOS provides a broader operating-system stack with Qernels, composable error mitigation, performance estimation, compatibility-based multiprogramming with buffer zones, and multi-objective scheduling across QPUs~\cite{giortamis2025qosquantumoperating}.

\name{} instead targets a forward-looking service-deployment model and a later system layer. It compiles stable workloads into reusable executable-version libraries during development. At each post-deployment invocation, it performs only lightweight orchestration. HyperQ and QOS decide which jobs co-execute and how they are scheduled, whereas \name{} assumes that the co-execution set has been selected and chooses one precompiled executable per workload. This choice determines per-invocation placement, avoids region conflicts, and favors high-fidelity versions. Thus, HyperQ and QOS optimize submitted-job execution, while \name{} avoids repeated compilation for a deployed service.

Ensembled quantum computing (EQC) uses multiple devices to execute circuits simultaneously, accelerating variational quantum algorithm training and counteracting system-specific biases~\cite{steinEQCEnsembledQuantum2021}; subsequent work profiles broader algorithms~\cite{khareParallelizingQuantumClassicalWorkloads2023}.
EQC exploits internal parallelism within one application across devices, whereas \name{} co-executes multiple applications on one device, making the approaches complementary.

Ensemble of Diverse Mappings (EDM)~\cite{tannuEnsembleDiverseMappings2019} mitigates correlated errors by combining program outputs from varied qubit allocations.
Although \name{} also compiles different mappings, it additionally targets online compilation latency, uses noise-aware partitioning to generate linearly scaling non-overlapping mappings rather than exponentially many isomorphic subgraphs, and combines compilation-runtime co-design with post-compilation fidelity ranking and runtime orchestration instead of acting as a standalone compiler.

\section{Discussion}
\label{sec:discussion}

\name{} is most effective when a quantum workload has a stable circuit skeleton that can be compiled during development and then invoked repeatedly after deployment.
This service-deployment model amortizes development-time compilation across post-deployment invocations but does not cover every workload.
QAOA-style programs may require regeneration when an input graph changes the circuit topology, although fixed-graph variants can reuse versions when only angles change.
Similarly, quantum neural networks may combine an input-dependent data-encoding front-end with a stable network structure; \name{} can amortize compilation of the reusable portion while regenerating the input-dependent portion.

The deployment model introduces storage and metadata overhead for multiple executable versions.
A version can be represented by a compact region-specific control binary and metadata for its target region, logical-to-physical mapping, parameter interface, and selection features rather than heavyweight compiler objects.
Candidate placements use compute-unit granularity, so the number of versions depends on feasible compute-unit combinations instead of all qubit-level subgraphs and does not grow combinatorially with total device size.
Nevertheless, large-scale library management, deduplication, and metadata indexing require further quantitative study.

The compile-time dominance in Figure~\ref{fig:compilation_vs_duration} is grounded in the studied superconducting-cloud setting and is not universal across hardware modalities.
Longer gate times on platforms such as trapped ions and neutral atoms can increase the QPU fraction of latency and weaken this conclusion~\cite{zhaoHighFidelityTwoQubitQuantum2025}, while recent neutral-atom systems already report sub-microsecond two-qubit gates~\cite{everedHighfidelityParallelEntangling2023}.
Compilation dominance is therefore hardware and time dependent.

The design is agnostic to device topology. Its compute-unit abstraction requires only a weighted connectivity graph and is not tied to heavy-hex. Regular or symmetric topologies may yield structurally similar regions, reducing the number of distinct patterns that need to be compiled and tracked. In contrast, irregular or calibration-heterogeneous devices may increase the value of maintaining region-specific versions.

The design deliberately limits its optimization boundary to post-selection runtime orchestration under a non-preemptive NISQ model.
After a co-execution set is selected, it chooses one conflict-free executable version per workload but does not decide which jobs co-execute, when they run, or how a queue scheduler prioritizes submissions.
It also omits online inter-program co-optimization requiring a joint view of all programs, including global routing, cross-program SWAP insertion, and joint remapping, and assumes neither long-term quantum memory nor context switching.
Although these capabilities could relax the state-preservation constraints on temporal sharing, they remain outside the studied model~\cite{xu2023systems_arch_qram,ding2020quantum}.
This boundary keeps the system focused on per-invocation placement for long-lived deployed services rather than a full queue scheduler or temporally shared quantum operating system.

\section{\texorpdfstring{Conclusion}{Conclusion}}

This paper presents \name{}, a system architecture that incorporates
(i) an architectural abstraction that models a quantum device as compute units,
(ii) a multi-version compiler with a lightweight cost function for executable ranking, and
(iii) a lightweight runtime with fidelity-aware orchestration.
By addressing the portability and fidelity challenges of offline compilation, the system removes online compilation from the critical execution path of MPQC designs.
Our evaluation on noisy simulators and real quantum devices shows that post-scheduling executable-version orchestration can reduce critical-path latency while maintaining comparable fidelity in conflict-free multiprogram execution.

\begin{acks}

We thank the anonymous reviewers of ASPLOS 2024, ISCA 2024, SOSP 2024, and HPCA 2025 for their valuable comments. We thank Mr. Ziming Qu for his help with the experiments.
Ying Wang is the corresponding author.
\end{acks}

\bibliographystyle{ACM-Reference-Format}
\bibliography{references}

\end{document}